\documentclass[12pt]{article}

\usepackage{hyperref}
\usepackage{cite}
\usepackage{units}
\usepackage{epsfig}
\usepackage{pifont}
\usepackage{amsmath,amsfonts,amssymb,amsthm,mathrsfs}
\usepackage{graphicx,chngcntr,slashed}
\usepackage{cancel}
\usepackage{pstricks}
\usepackage{afterpage}
\usepackage{braket}
\usepackage{caption}
\usepackage{subcaption}
\usepackage{xcolor}
\usepackage{makecell}
\usepackage{physics}
\usepackage{multirow}
\usepackage{breqn}

\newcommand{\be}{\begin{equation}}
\newcommand{\ee}{\end{equation}}
\newcommand{\bea}{\begin{eqnarray}}
\newcommand{\eea}{\end{eqnarray}}

\makeatletter
\def\section{\@startsection {section}{1}{\z@}{+3.0ex plus +1ex minus
  +.2ex}{2.3ex plus .2ex}{\large\bf\boldmath}}
\def\subsection{\@startsection{subsection}{2}{\z@}{+2.5ex plus +1ex
minus +.2ex}{1.5ex plus .2ex}{\normalsize\bf\boldmath}}
\def\subsubsection{\@startsection{subsubsection}{3}{\z@}{+3.25ex plus
 +1ex minus +.2ex}{1.5ex plus .2ex}{\normalsize\it}}

\DeclareCaptionFormat{cont}{#1 (cont.)#2#3\par}
\graphicspath{{.}{plots/}}

\begin{document}

\def\thefootnote{\fnsymbol{footnote}}

\begin{flushright}
\end{flushright}

\vspace{1cm}

\begin{center}

{\Large {\bf Renormalization-group running of dimension-8 four-fermion operators in the SMEFT}}
\\[3.5em]
{\large
Radja~Boughezal$^1$, Yingsheng Huang$^{1,2}$ and Frank~Petriello$^{1,2}$ 
}

\vspace*{1cm}

{\sl
$^1$ HEP Division, Argonne National Laboratory, Argonne, Illinois 60439, USA \\[1ex]
$^2$ Department of Physics \& Astronomy, Northwestern University,\\ Evanston, Illinois 60208, USA
}

\end{center}

\vspace*{0.5cm}

\begin{abstract}
    
We compute the renormalization-group equations governing the evolution of dimension-8 four-fermion operators in the Standard Model Effective Field Theory (SMEFT). We describe the calculation and present analytic results for both the full flavor structure of the SMEFT and with the assumption of minimal flavor violation. We present numerical results for the renormalization-group evolution of the coefficients, and study their impact on fits of the Large Hadron Collider (LHC) Drell-Yan data. The effects of running on the dimension-8 coefficients can reach 50\% or more when evolving from 10 TeV scale down to few-GeV energies relevant for the analysis of fixed-target data. However, the impact of the dimension-8 running on the analysis of Drell-Yan data from the LHC is minimal.
    
\end{abstract}



\section{Introduction}

The Standard Model (SM) of particle physics successfully describes processes ranging from low-energy nuclear phenomena to high-energy collisions. However, since it does not contain dark matter, and cannot explain certain observations such as the matter-antimatter asymmetry in the universe, we expect that undiscovered physics beyond the SM that explains these mysteries exists. Experiments at the Large Hadron Collider (LHC) and elsewhere are probing the TeV scale, searching for solutions to these outstanding problems. No conclusive deviation from SM predictions has yet been found despite probes now reaching multi-TeV levels in many channels, suggesting that a mass gap exists between any new physics and the electroweak scale. A major theme of current research is to understand how heavy new physics can be indirectly probed by available and upcoming data. 

A convenient theoretical framework for investigating indirect signatures of heavy new physics is the SM Effective Field Theory (SMEFT). The SMEFT is formed by adding higher-dimensional operators to the SM Lagrangian that are consistent with the SM gauge symmetries and formed only from SM fields. The higher-dimensional operators in the SMEFT are suppressed by appropriate powers of a high energy scale $\Lambda$ below which heavy new fields are integrated out. The SMEFT encapsulates a broad swath of new physics models, making it easier to simultaneously study numerous theories without focusing on details of their ultraviolet completions that do not matter at low energies. Complete, non-redundant bases for the dimension-6~\cite{Buchmuller:1985jz,Arzt:1994gp,Grzadkowski:2010es} and dimension-8 operators~\cite{Murphy:2020rsh,Li:2020gnx} in the SMEFT have been constructed, and the full basis of operators has recently been determined through dimension-12~\cite{Harlander:2023psl}.

One key aspect in applications of the SMEFT to the analysis of experimental data is the need for global fits of the available data. The number of additional operators introduced even at the dimension-6 level requires a diverse set of data to constrain all directions in parameter space. There has been significant recent work devoted to performing global fits to the available LHC data~\cite{Pomarol:2013zra,DiVita:2017eyz,Almeida:2018cld,Ellis:2018gqa,Cirigliano:2019vfc,Hartland:2019bjb,Brivio:2019ius,vanBeek:2019evb,Aoude:2020dwv,Ellis:2020unq,Dawson:2020oco,Greljo:2021kvv,Ethier:2021bye,Aoude:2022aro,Cirigliano:2023nol,Costantini:2024xae}. Certain Wilson coefficients in the SMEFT are poorly probed at the LHC due to flat directions in parameter space, and are better measured at lower-energy experiments such as the EIC~\cite{Boughezal:2020uwq,Boughezal:2022pmb} or fixed-target parity-violating electron scattering~\cite{Boughezal:2021kla,Crivellin:2021bkd,Wang:2024mll}.

A second important aspect of ongoing SMEFT analyses is the precision of the theoretical predictions used. There has been much recent effort in computing the necessary SMEFT predictions to higher orders in the coupling constants. Tools are available for the computation of QCD corrections to 
arbitrary processes within the SMEFT~\cite{Degrande:2020evl}. For processes such as Drell-Yan production of lepton pairs the full one-loop corrections in the SMEFT exist~\cite{Dawson:2018dxp,Dawson:2021ofa}. Given the need motivated above to incorporate experiments at different energy scales into global fits one important component of the higher-order corrections are the renormalization-group (RG) evolution of the SMEFT Wilson coefficients. The full one-loop running in SMEFT is known at the dimension-6 level~\cite{Jenkins:2013zja,Jenkins:2013wua,Alonso:2013hga}, and efforts toward extending this to higher loop orders are ongoing~\cite{Bern:2020ikv,Alonso:2022ffe,Jenkins:2023bls,Banerjee:2024rbc}. 
{Apart from the conventional Feynman-diagram approach, alternative methods have been proposed to advance the computation of SMEFT RG equations (RGEs), such as on-shell methods~\cite{Bern:2020ikv,Jiang:2020mhe,AccettulliHuber:2021uoa}, geoSMEFT~\cite{Helset:2022pde,Assi:2023zid}, functional matching~\cite{Henning:2016lyp,Zhang:2016pja,Cohen:2020fcu,Born:2024mgz}, and the heat kernel method~\cite{Jack:1982hf,Buchalla:2019wsc,Banerjee:2023iiv,Chakrabortty:2023yke,Banerjee:2023xak}. }
The impact of the RG evolution in fits to LHC data has been the focus of recent interest~\cite{Maltoni:2024dpn} {and} the interplay between RG evolution and lepton-flavor-violating experiments has also been investigated~\cite{Ardu:2021koz,Ardu:2022pzk}. 

The SMEFT involves an expansion in higher-dimensional operators as well as in the coupling constants. Dimension-8 terms can have important phenomenological impact, for example at LHC energies in the Drell-Yan process~\cite{Alioli:2020kez,Boughezal:2021tih,Boughezal:2023nhe}, and in precision electroweak studies at lower energies
~{\cite{Banerjee:2023qbg}}
. The effect of the dimension-8 terms in fits depend strongly upon the underlying ultraviolet (UV) model~\cite{Dawson:2022cmu,Ellis:2023zim,Dawson:2024ozw}, In some instances they can serve as diagnostic tools to distinguish between different potential UV completions of the effective theory~\cite{Boughezal:2022nof}. The possible ranges of Wilson coefficients at dimension-8 obey positivity constraints that can be tested with the available data~\cite{Li:2022rag,Chen:2023bhu}.

Given the potential impact of dimension-8 operators, the need for precision theoretical predictions, and the fact that global fits necessarily involve experiments at different scales in order to probe all directions in the Wilson coefficient parameter space, it is important to extend the calculation of the RG equations (RGEs) of the SMEFT operators to the dimension-8 level. We pursue that goal in this work by studying the RGEs of the four-fermion sector of the SMEFT at dimension-8. We focus on the Hermitian operators formed from the product of two vector currents that appear in numerous new physics models. We derive the RG equations, both in full generality and when restricted to the minimal flavor violation (MFV) flavor structure, an assumption often made in fits to the experimental data. Important initial work on the RG evolution of the SMEFT at dimension-8 was done in~\cite{Chala:2021pll,Chala:2021wpj,DasBakshi:2022mwk,Chala:2023jyx,Chala:2023xjy,DasBakshi:2023htx}, where the focus was on whether the RG running preserves the tree-level positivity constraints.  Our main interest here is in the phenomenological impact of the RGEs, and we study the dimension-8 RGE impact both on the evolution of the associated Wilson coefficients and on fits to the Drell-Yan data at the LHC. We show that the running of the coefficients at the dimension-8 level can reach tens of percent when running from high scales above the LHC reach to the scales relevant for the analysis of Drell-Yan data. The evolution can reach over 50\% when running to scales relevant for low-energy experiments. We study fits to the high-energy Drell-Yan invariant mass and forward-backward asymmetry data from the LHC in the context of matching explicit ultraviolet completions to the SMEFT.  In general the RG running at both dimension-6 and dimension-8 has a small effect on bounds at the LHC obtained from the Drell-Yan process, suggesting that at least in this sector of the SMEFT that the effect of higher-order loop corrections to the dimension-8 terms are under good theoretical control.

Our paper is organized as follows. We briefly review the SMEFT and establish our notation in Section~\ref{sec:notation}. In Section~\ref{sec:calc} we discuss our calculation of the dimension-8 RGEs. We discuss the techniques that we use to disentangle the operator mixing at the dimension-8 level, and we discuss the simplifications of the RGEs that occur when assuming minimal flavor violation. We present our numerical results for the running of the Wilson coefficients and on the impact of the RG running on fits to LHC Drell-Yan data in Section~\ref{sec:numerics}. We conclude in Section~\ref{sec:conc}. In the Appendix we present the dimension-8 RGEs assuming minimal flavor violation.

\section{Notations and conventions}
\label{sec:notation}

We begin with a brief review of the SMEFT, focusing on aspects relevant for our analysis.  In our study we keep terms through dimension-8 in the $1/\Lambda$ expansion and 
ignore operators of odd dimension which violate lepton number. With these assumptions the SMEFT Lagrangian becomes
\begin{equation}
{\cal L} = {\cal L}_{SM}+ \frac{1}{\Lambda^2} \sum_i C_{i}^{(6)} {\cal
  O}_{i}^{(6)} + \frac{1}{\Lambda^4} \sum_i C_{i}^{(8)} {\cal
  O}_{i}^{(8)} + \ldots,
\end{equation}
where the ellipsis denotes operators of higher dimensions.  The Wilson
coefficients $C_{i}^{(6)}$ and $C_{i}^{(8)}$ defined above are dimensionless. 
Cross sections computed through ${\cal O}(1/\Lambda^4)$ will have contributions 
from the square of dimension-6 operators, as well as interferences between dimension-8 operators and the SM. In our study we focus on four-fermion operators. These play a central role in new physics searches in the Drell-Yan channel at the LHC, as well as in the deep-inelastic scattering process at the EIC and in lower-energy fixed-target experiments, and appear ubiquitously in models containing extra $Z^{\prime}$ bosons arising from additional U(1) gauge groups in the UV. A complete listing of the operators relevant for our analysis is given in Table~\ref{tab:processes}. We note that the four-fermion operators considered here mix with other operators under RG evolution that contain Higgs fields, such as ${\cal O}_{\phi f} = (H^{\dagger} i \overset{\leftrightarrow}{D}_{\mu} H) (\bar{f} \gamma^{\mu} f)$ at dimension-6, where $f$ denotes a SM fermion field, and analogous operators at dimemsion-8. This can be seen from the dimension-6 RGEs in~\cite{Jenkins:2013zja,Jenkins:2013wua,Alonso:2013hga}, for example. In explicit $Z^{\prime}$ models these operators are generated only by kinetic mixing and the Higgs contribution to the $Z^{\prime}$ mass that arises from the Lagrangian term
\be
Z^{\prime}_{\mu}Z^{\prime \mu} |H^{\dagger} H| ,
\ee
as can be seen from~\cite{Dawson:2024ozw}. As a first step we set these terms to zero, which assumes they are small compared to the terms we study, to simplify this initial analysis of the dim-8 RGEs which is meant to determine whether dim-8 RG running has a significant impact on experimental analyses. In the SMEFT this means we focus only on the four-fermion sector and neglect mixing with these other operators.

The purpose of our work is to derive the renormalization-group equations of a subset of the dimension-8 Wilson coefficients consisting of four-fermion operators formed from a product of two vector currents. To do so we write the structure of the RG equations as
\begin{align}
	&\dv{C_i^{(6)}}{\ln\mu}=\sum_j\gamma_{ij}^{(6)}C_j^{(6)},\\
	&\dv{C_i^{(8)}}{\ln\mu}=\sum_j\gamma_{ij}^{(8)}C_j^{(8)}+\sum_{j,k}\gamma_{ijk}^{(6)}C_j^{(6)}C_k^{(6)},
\end{align}
where $\mu$ denotes the SMEFT renormalization scale. The matrices $\gamma_{ij}^{(6)}$, $\gamma_{ij}^{(8)}$ and $\gamma_{ijk}^{(6)}$ have off-diagonal entries, indicating mixing between the operators. We note that the dimension-8 Wilson coefficients do not affect the dimension-6 running, as they first appear at higher order in the $1/\Lambda^2$ expansion. The dimension-6 terms do lead to running of the dimension-8 Wilson coefficients, and to a non-linear term in the system of differential equations. In the next sections we use the following notation in order to compare with~\cite{Alonso:2013hga}:
\begin{align}
	\dot{C}\equiv16\pi^2\mu\frac{\mathrm{d}}{\mathrm{d}\mu}C .
\end{align}
We renormalize the SMEFT multiplicatively using standard $\overline{{\rm MS}}$ subtraction at the one-loop order by introducing $Z$-factors for each operator and field-strength renormalizations for the fields in each operator. We write the Lagrangian in terms of renormalized Wilson coefficients and operators as
\begin{equation}
\mathcal{L} = \sum_{i,j} Z_{ij} \sqrt{Z_i^{wf}} C_j \mathcal{O}_i,
\label{eq:Zdef}
\end{equation}
where
\begin{equation}
\sqrt{Z_i^{wf}}  = \Pi_a \sqrt{Z_{i,a}^{wf}}
\end{equation}
denotes a product over wave function renormalization factors for each field $a$ in the operator $\mathcal{O}_i$. We expand the $Z$ factors perturbatively around unity. We note for future reference the following values of the wave-function renormalization constants, which can be derived from a simple one-loop calculation in the SM
\begin{subequations}
	\begin{align}
		&\delta_{q}=\frac{g_s^2C_F}{16\pi^2\epsilon}+\frac{g_1^2+27g_W^2}{576\pi^2\epsilon},\\ 
		&\delta_{u}=\frac{g_s^2C_F}{16\pi^2\epsilon}+\frac{g_1^2}{36\pi^2\epsilon},\\ 
		&\delta_{d}=\frac{g_s^2C_F}{16\pi^2\epsilon}+\frac{g_1^2}{144\pi^2\epsilon},\\ 
		&\delta_{l}=\frac{g_1^2+3g_W^2}{64\pi^2\epsilon},\\ 
		&\delta_{e}=\frac{g_1^2}{16\pi^2\epsilon}
	\end{align}
\end{subequations}
where $q$ denotes the left-handed quark, $u$ denotes the right-handed up-type quark, $d$ denotes the right-handed down-type quark, $l$ denotes the left-handed lepton, and $e$ denotes the right-handed charged lepton. $g_1$ denotes the hypercharge gauge coupling, $g_W$ denotes the SU(2) gauge coupling, $g_s$ denotes the SU(3) strong coupling constant, $C_F=4/3$ is the usual SU(3) Casimir constant, and we work in $d=4-2\epsilon$ space-time dimensions. We have expanded the wave-function renormalization factors as $Z_i^{wf}= 1-\delta_i$. We note that the dim-6 four-fermion contributions to the wave-function renormalization lead to scaleless tadpole integrals that can consistently be set to zero.

\section{Calculation of the renormalization group equations}
\label{sec:calc}

We list all dimension-6 and dimension-8 operators relevant for our analysis in the left-hand column of Table~\ref{tab:processes}. Our notation for these operators follows that in~\cite{Murphy:2020rsh}. There are 20 dimension-6 operators and 40 dimension-8 operators that enter this analysis. As mentioned in the previous section we focus on the four-fermion operators, neglecting operators that arise in completions of the SMEFT from kinetic mixing or Higgs-induced $Z^{\prime}$ mass contributions. We extract the counterterms from the one-loop on-shell amplitudes. Generically, a single process receives divergent one-loop corrections from multiple operators, leading to operator mixing. By utilizing the spinor chains, color factors, and kinematic structures that appear at tree level, we can identify and isolate the contributions of different operators to the one-loop UV poles. We outline below the specific tactics we use to accomplish this separation.
\begin{enumerate}

	\item When the operator is uniquely determined by the chirality of the external fermions, the UV pole is straightforwardly associated with a single operator at each order in $\Lambda$. For example, the electron-positron scattering process with right-handed external states $e_R^+ e_R^- \to e_R^+ e_R^-$ receives contributions only from ${\cal O}_{ee}$ at dimension-6 and ${\cal O}_{e^2 D^4}$ at dimension-8, making it simple to determine the renormalization constants associated with these terms. This assumes linear insertions of these operators; the quadratic insertions of dimension-6 operators are straightforward to disentangle.
	
	\item At each order in $\Lambda$ there are multiple $(\bar LL)(\bar LL)$-type operators with and without an $\mathrm{SU}(2)_L$ generator $\tau$. We use the sums and differences of amplitudes for two processes to separate their renormalization constants. For example, at dimension-6 both $\mathcal{O}_{lq}^{(1)}$ and $\mathcal{O}_{lq}^{(3)}$ contribute to $u_L\bar u_L\to e_L^+e_L^-$ and $d_L\bar d_L\to e_L^+e_L^-$ at tree level. We can express the LO amplitude for each process as
\begin{subequations}
		\begin{align}
			\mathcal{M}^{0}(u_L\bar u_L\to e_L^-e_L^+)=-\frac{C_{lq}^{(1)}-C_{lq}^{(3)}}{\Lambda^2}\delta_{c_1c_2}\;\bar v_u(k_2)\gamma^\mu P_L u_u(k_1)\;\bar u_e(k_3)\gamma_\mu P_L v_e(k_4)\\
			\mathcal{M}^{0}(d_L\bar d_L\to e_L^-e_L^+)=-\frac{C_{lq}^{(1)}+C_{lq}^{(3)}}{\Lambda^2}\delta_{c_1c_2}\;\bar v_d(k_2)\gamma^\mu P_L u_d(k_1)\;\bar u_e(k_3)\gamma_\mu P_L v_e(k_4)
		\end{align}
\end{subequations}
where $P_L$ is the left-handed chirality projector, and $p_i$ and $c_i$ are the momentum and color index of the $i$-th particle. We neglect here the flavor indices for notational simplicity. We note that the sum of these two processes only depends on $C_{lq}^{(1)}$, while the difference only depends on $C_{lq}^{(3)}$. We can therefore use the sum and difference of these two processes to extract the counterterms for $C_{lq}^{(1)}$ and $C_{lq}^{(3)}$, respectively. As an example we consider the contribution of the Wilson coefficient $C_{qe}$ to the one-loop UV pole for each of these processes. The relevant amplitudes take the form
\begin{eqnarray}
		\mathcal{M}^{1,C_{qe}}_{\mathrm{UV}}(u_L\bar u_L\to e_L^-e_L^+)
		 &=& \mathcal{M}^{1,C_{qe}}_{\mathrm{UV}}(d_L\bar d_L\to e_L^-e_L^+) \nonumber \\ 
		 &=&\frac{g_1C_{qe}}{48\pi^2\epsilon\Lambda^2}\delta_{c_1c_2}\;\bar v_{u/d}(k_2)\gamma^\mu P_L u_{u/d}(k_1)\;\bar u_e(k_3)\gamma_\mu P_L v_e(k_4). \nonumber \\
\end{eqnarray}
The difference of these two one-loop amplitudes vanishes, indicating that $C_{qe}$ does not contribute to the one-loop running of $C_{lq}^{(3)}$. This allows us to isolate the contribution of $C_{qe}$ to the running of $C_{lq}^{(1)}$. We employ the same technique at dimension-8 to separate operators with and without $\tau$. The sums and differences used are indicated explicitly in the second column of 
Table~\ref{tab:processes}.

	\item Some operators contain the $\mathrm{SU}(3)_c$ generator $T$. They contribute to the same process as their counterparts without $T$. We use the distinct color factors associated with each operator to disentangle them. For example, $\mathcal{O}_{ud}^{(1)}$ and $\mathcal{O}_{ud}^{(8)}$ both contribute to $u_R\bar u_R\to d_R\bar d_R$. The former corresponds to a color factor of $\delta_{pr}\delta_{st}$, while the latter corresponds to $T^A_{pr}T^A_{st}$. The tree-level amplitude for the process that demonstrates this is shown below:
\begin{align}
		\mathcal{M}^{0}(u_R^p\bar u_R^r\to d_R^s\bar d_R^t)=&-\frac{C_{ud}^{(1)}\delta_{pr}\delta_{st}+C_{ud}^{(8)}T^A_{pr}T^A_{st}}{\Lambda^2}\bar v_u(k_2)\gamma^\mu P_R u_u(k_1)\;\bar u_d(k_4)\gamma_\mu P_R v_d(k_3),
\end{align}
where $P_R$ is the right-handed chirality projector, and $p$, $r$, $s$, $t$ are the color indices of the external particles. These unique color factors are then utilized to extract the counterterms for $C_{ud}^{(1)}$ and $C_{ud}^{(8)}$.  	
	
	\item To disentangle effects of dimension-8 operators that contribute to the same process, we also exploit their distinct dependence on kinematic structures. For example, $C_{l^4D^2}^{(1)}$ and $C_{l^4D^2}^{(2)}$ both contribute to $\nu\bar\nu\to e_L^+e_L^-$, but the former is proportional to $t+u$ at tree-level while the latter is proportional to $t-u$, where $s$, $t$ and $u$ are the standard Mandelstam variables. The terms proportional to $t+u$ and $t-u$ in the one-loop UV poles contribute to the renormalization of $C_{l^4D^2}^{(1)}$ and $C_{l^4D^2}^{(2)}$, respectively. 
\end{enumerate}

\begingroup 
\renewcommand{\arraystretch}{1.5}
\begin{table}
	\centering
	\begin{tabular}[H]{cccc}
		\hline\hline
		&Operator & Process\\
		\hline 
		\multirow{5}{*}{$(\bar LL)(\bar LL)$}
		&$\mathcal{O}_{ll}, \mathcal{O}_{l^4D^2}^{(1)}, \mathcal{O}_{l^4D^2}^{(2)}$ & $\nu\bar\nu\to e_L^+e_L^-$\\
		&$\mathcal{O}_{qq}^{(1)}, \mathcal{O}_{q^4D^2}^{(1)}, \mathcal{O}_{q^4D^2}^{(1)}$ & $\pqty{u_L\bar u_L\to u_L\bar u_L}+\pqty{u_L\bar u_L\to d_L\bar d_L}$\\
		&$\mathcal{O}_{qq}^{(3)}, \mathcal{O}_{q^4D^2}^{(3)}, \mathcal{O}_{q^4D^2}^{(4)}$ & $\pqty{u_L\bar u_L\to u_L\bar u_L}-\pqty{u_L\bar u_L\to d_L\bar d_L}$\\
		&$\mathcal{O}_{lq}^{(1)}, \mathcal{O}_{l^2q^2D^2}^{(1)}, \mathcal{O}_{l^2q^2D^2}^{(1)}$ & $\pqty{u_L\bar u_L\to e_L^+e_L^-}+\pqty{d_L\bar d_L\to e_L^+e_L^-}$\\
		&$\mathcal{O}_{lq}^{(3)}, \mathcal{O}_{l^2q^2D^2}^{(3)}, \mathcal{O}_{l^2q^2D^2}^{(4)}$ & $\pqty{u_L\bar u_L\to e_L^+e_L^-}-\pqty{d_L\bar d_L\to e_L^+e_L^-}$\\
		\hline 
		\multirow{7}{*}{$(\bar RR)(\bar RR)$}
		&$\mathcal{O}_{ee}, \mathcal{O}_{e^4D^2}$ & $e_R^+e_R^-\to e_R^+e_R^-$\\
		&$\mathcal{O}_{uu}, \mathcal{O}_{u^4D^2}^{(1)}, \mathcal{O}_{u^4D^2}^{(2)}$ & $u_R\bar u_R\to u_R\bar u_R$\\
		&$\mathcal{O}_{dd}, \mathcal{O}_{d^4D^2}^{(1)}, \mathcal{O}_{d^4D^2}^{(2)}$ & $d_R\bar d_R\to d_R\bar d_R$\\
		&$\mathcal{O}_{eu}, \mathcal{O}_{e^2u^2D^2}^{(1)}, \mathcal{O}_{e^2u^2D^2}^{(2)}$ & $e_R^+e_R^-\to u_R\bar u_R$\\
		&$\mathcal{O}_{ed}, \mathcal{O}_{e^2d^2D^2}^{(1)}, \mathcal{O}_{e^2d^2D^2}^{(2)}$ & $e_R^+e_R^-\to d_R\bar d_R$\\
		&$\mathcal{O}_{ud}^{(1)}, \mathcal{O}_{u^2d^2D^2}^{(1)}, \mathcal{O}_{u^2d^2D^2}^{(2)}$ & $u_R\bar u_R\to d_R\bar d_R$\\
		&$\mathcal{O}_{ud}^{(8)}, \mathcal{O}_{u^2d^2D^2}^{(3)}, \mathcal{O}_{u^2d^2D^2}^{(4)}$ & $u_R\bar u_R\to d_R\bar d_R$\\
		\hline
		\multirow{8}{*}{$(\bar LL)(\bar RR)$}
		&$\mathcal{O}_{le}, \mathcal{O}_{l^2e^2D^2}^{(1)}, \mathcal{O}_{l^2e^2D^2}^{(2)}$ & $\nu\bar\nu\to e_R^+e_R^-$\\
		&$\mathcal{O}_{lu}, \mathcal{O}_{l^2u^2D^2}^{(1)}, \mathcal{O}_{l^2u^2D^2}^{(2)}$ & $u_R\bar u_R\to e_L^+e_L^-$\\
		&$\mathcal{O}_{ld}, \mathcal{O}_{l^2d^2D^2}^{(1)}, \mathcal{O}_{l^2d^2D^2}^{(2)}$ & $d_R\bar d_R\to e_L^+e_L^-$\\
		&$\mathcal{O}_{qe}, \mathcal{O}_{q^2e^2D^2}^{(1)}, \mathcal{O}_{q^2e^2D^2}^{(2)}$ & $u_L\bar u_L\to e_R^+e_R^-$\\
		&$\mathcal{O}_{qu}^{(1)}, \mathcal{O}_{q^2u^2D^2}^{(1)}, \mathcal{O}_{q^2u^2D^2}^{(2)}$ & $u_L\bar u_L\to u_R\bar u_R$\\
		&$\mathcal{O}_{qu}^{(8)}, \mathcal{O}_{q^2u^2D^2}^{(3)}, \mathcal{O}_{q^2u^2D^2}^{(4)}$ & $u_L\bar u_L\to u_R\bar u_R$\\
		&$\mathcal{O}_{qd}^{(1)}, \mathcal{O}_{q^2d^2D^2}^{(1)}, \mathcal{O}_{q^2d^2D^2}^{(2)}$ & $d_L\bar d_L\to d_R\bar d_R$\\
		&$\mathcal{O}_{qd}^{(8)}, \mathcal{O}_{q^2d^2D^2}^{(3)}, \mathcal{O}_{q^2d^2D^2}^{(4)}$ & $d_L\bar d_L\to d_R\bar d_R$\\
		\hline\hline
	\end{tabular}
	\caption{Processes used to extract the counterterms of individual operators. $\nu$ denotes the chiral left-handed neutrinos, $e$ denotes the charged leptons, $u$ denotes the up-type quarks and 
	$d$ denotes the down-type quarks. The chiralities of the external states are specified by the subscripts $L$ and $R$. 
	The sums and differences of processes are used to disentangle the contributions of operators with and without the $\mathrm{SU}(2)_L$ generator $\tau$, as explained in the text.
	\label{tab:processes}}
\end{table}
\endgroup

Once the counterterms are determined, we then subtract the field-strength renormalization constants from them to obtain the renormalization constants $Z_{ij}$ defined in Eq.~(\ref{eq:Zdef}). The RGEs are then readily read off from these quantities
\begin{align}
	\dv{C_{\mathcal{O}}}{\ln\mu}=-\dv{Z_{\mathcal{O}}}{\ln\mu}.
\end{align}
As a check we have applied the procedure outlined above to recompute the dimension-6 RGEs and found complete agreement with results in the literature~\cite{Jenkins:2013zja,Jenkins:2013wua,Alonso:2013hga}. The full dimension-8 RGEs are included in attached Mathematica files with this submission.

\subsection{RGEs in minimal flavor violation}

An often-used assumption in SMEFT analyses is that of minimal flavor violation (MFV). This simplifies the flavor structure of the Wilson coefficients and reduces the number of free parameters in global analyses while being in good agreement with experimental data. We present our results for the RGEs assuming MFV in the Appendix, and give here a brief overview of the pertinent details. A more complete description of the implications of MFV for SMEFT fits can be found in~\cite{Aoude:2020dwv}.

The defining assumption of MFV is that the SMEFT Lagrangian is invariant under the following $U(3)^5$ flavor symmetry broken only by the fermion Yukawa couplings:
\be
U(3)_q \times U(3)_u \times U(3)_d \times U(3)_l \times U(3)_e.
\ee
The SM fermions are charged as follows under this symmetry:
\be
q \sim (3,1,1,1,1),\;\; u\sim (1,3,1,1,1), \;\; d\sim (1,1,3,1,1), \;\; l\sim (1,1,1,3,1), \;\; e\sim (1,1,1,1,3).
\ee
Barred fermions transform under the $\bar{3}$ representation of the appropriate symmetry group. The Yukawa matrices are treated as spurions under this symmetry transformation. To understand the implications of MFV for the structure of the RGEs we consider here the transformation of several example four-fermion operators under the above symmetry. We first consider the semileptonic operator
\be
{\cal O}_{lq} = (\bar{q}_p \gamma^{\mu} q_r ) (\bar{l}_s \gamma_{\mu} l_t )
\ee
where the subscripts denote the fermion flavor indices. To make this invariant under the $U(3)^5$ symmetry we must contract the lepton-doublet and quark-doublet indices, indicating that the Wilson coefficient has the structure
\be
C_{prst} = C \delta_{pr} \delta_{st}.
\ee
Only a single structure is allowed for this Wilson coefficient under the MFV assumption. We now consider the operator
\be
{\cal O}^{(1)}_{qq} = (\bar{q}_p \gamma^{\mu} q_r ) (\bar{q}_s \gamma_{\mu} q_t )
\ee
In this case there are two possible contractions invariant under the $U(3)^5$ symmetry, and the Wilson coefficient of this operator takes the form
\be
C_{prst} = C \delta_{pr} \delta_{st}+C^{\prime} \delta_{pt} \delta_{sr}.
\ee
Two independent structures are allowed. The following operators have two independent Wilson coefficient structures under the MFV assumption.
\begin{itemize}

\item left-handed operators: $\mathcal{O}_{ll}$, $\mathcal{O}_{qq}^{(1)}$ and $\mathcal{O}_{qq}^{(3)}$, as well as their dimension-8 counterparts $\mathcal{O}_{l^4D^2}^{(1-2)}$ and $\mathcal{O}_{q^4D^2}^{(1-4)}$; 

\item right-handed operators: $\mathcal{O}_{uu}$, $\mathcal{O}_{dd}$, as well as their dimension-8 counterparts $\mathcal{O}_{u^4D^2}^{(1-2)}$ and $\mathcal{O}_{d^4D^2}^{(1-2)}$. 

\end{itemize}
All other four-fermion operators contain only a single structure. The use of MFV simplifies the computation of the RGEs. By selecting two sets of specific generation indices we can reconstruct the separate RG evolution of $C$ and $C^{\prime}$. For example choosing $p=r=1$ and $s=t=2$ isolates the evolution of $C$, while selecting $p=t=1$ and $s=r=2$ isolates the $C^{\prime}$ contribution. We show the RGEs of the dimension-8 Wilson coefficients assuming MFV in the Appendix.

\section{Numerical results}
\label{sec:numerics}

Our primary interest in the RGEs is their impact on the analysis of experimental data. We present here numerical results that illustrate these effects. We solve the RG equations numerically as they contain non-linear contributions at the dimension-8 order. We first show the running of selected dim-6 and dim-8 Wilson coefficients. We then consider an analysis that illustrates the impact of RG running on experimental analyses. We assume several UV models containing $Z^{\prime}$ gauge bosons, match these models to the SMEFT, and RG evolve the Wilson coefficients from the matching scale to the relevant scale of the experimental analysis. Our focus in this example is on the Drell-Yan process at the LHC, and we consider fits to existing 13 TeV invariant mass and forward-backward asymmetry data and study the impact of running on the UV model-parameter bounds. This illustrate the running effect on fits at the LHC, and also shows the impact of running when using the SMEFT to extract the parameters of an explicit UV model. Since one of our primary interests in deriving the RGEs is to investigate the convergence of the SMEFT expansion we perform this analysis with several assumptions: truncating the SMEFT expansion at $1/\Lambda^2$ without RG running, truncating the SMEFT expansion at $1/\Lambda^2$ and including RG running, keeping the  $1/\Lambda^4$ terms without running, and finally keeping the  $1/\Lambda^4$ terms and including the RG running. This allows us to investigate how each of these contributions to the expansion affects the analysis of LHC data.

\begin{table}[pt]
	\centering
	\begin{tabular}[htbp]{|c||c|c|c|}
		\hline\hline
		SM fermion & $Y_D$ & $B_L$ & inert $E_6$ \\
		\hline\hline
		q & 1/6 & 1/3 & 0 \\
		u & 2/3 & 1/3 &  0\\
		d & -1/3 & 1/3 & -1/2\\
		l & -1/2 & -1 & -1/2 \\
		e & -1 & -1 & 0 \\
		\hline\hline
\end{tabular}
	\caption{Charges of the SM fermions under the $Z^{\prime}$ gauge groups considered in our analysis.
		\label{tab:Zpmodels}}
\end{table}
\subsection{Running of the Wilson coefficients}

We first illustrate the size of the RG running by matching several UV models to the SMEFT at the scale $\Lambda = M_{Z^{\prime}}$ which we assume to be 10 TeV, and running the RGEs to a lower scale $\mu$ for several representative coefficients at both dimension-6 and dimension-8. We perform this matching at tree-level. The example UV models contain $Z^{\prime}$ bosons and are motivated by recent work in~\cite{Dawson:2024ozw}.We use a subset of these $Z^{\prime}$ models, specifically the $Y_D$, $B_L$, and inert $E_6$ models~\footnote{{The $E_6$ model here serves solely as an example of an anomaly-free UV model featuring additional U(1) symmetries, with no implications for the underlying grand unified theory.}}, which illustrate the important aspects of the running. The charges of the SM fermions in these models are shown in Table~\ref{tab:Zpmodels}. We note that since we assume the Higgs contribution to the $Z^{\prime}$ mass is small compared to other effects, as discussed in an earlier section, the physical mass of the $Z^{\prime}$ boson and its Lagrangian parameter coincide. Plots for representative couplings are shown in Fig.~\ref{fig:running}. We see that the effects of the running of the dimension-6 operators range from 5-10\% when running from 10 TeV to 100 GeV, with the effect growing to 1{0}-20\% when running to 10 GeV, as is relevant for lower-energy experiments. The running effects on the dimension-8 coefficients reach 40-{8}0\% when running to 10 GeV. This is simply due to  larger numerical coefficients in the RG equations at dimension-8 than at dimension-6. We note that many coefficients, such as the $C^{\prime}$ coefficients in the context of MFV, start at zero with tree-level matching but are generated at lower scales by RG running. These generically remain small even upon running to lower scales. We note that the results are very dependent on the specific $Z^{\prime}$ model chosen, with much larger effects being obtained for the $Y_D$ and $B_L$ models than for the inert $E_6$ model. This is because, as shown in Table~\ref{tab:Zpmodels}, the first two models turn on more non-zero coefficients at the starting scale $\mu_0$, which leads to more contributions to the running. In particular we note that the quadratic dimension-6 contributions to the RG evolution have a large impact on the evolution of the dimension-8 terms.

\begingroup
\makeatletter
\providecommand\phantomcaption{\caption@refstepcounter\@captype}
\makeatother
\newcommand{\subfigwidth}{0.49\linewidth}
\newcommand{\figcap}{
	\caption{Running of representative Wilson coefficients in the $Z^{\prime}$ models defined in the text. The initial scale $\mu_0$ is set to the new physics scale $\Lambda=M_{Z'}=10\ \mathrm{TeV}$. The lower inset in each plot shows the relative deviation from the value at the starting scale.}}
\begin{figure}[!htbp]
	\centering
	{\includegraphics[width=\subfigwidth]{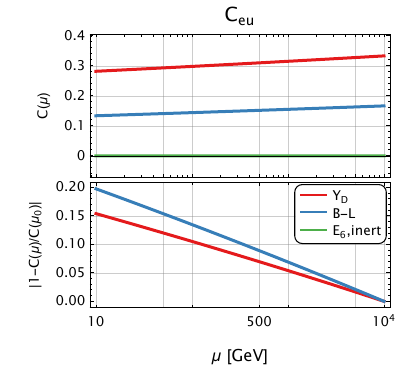}}\hfill
	{\includegraphics[width=\subfigwidth]{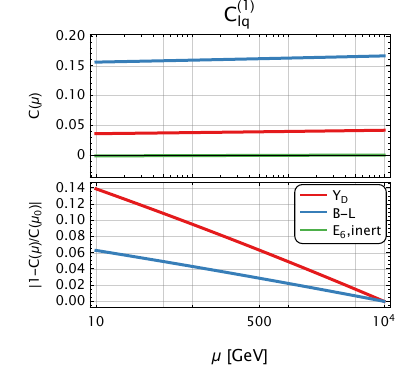}}\\
	{\includegraphics[width=\subfigwidth]{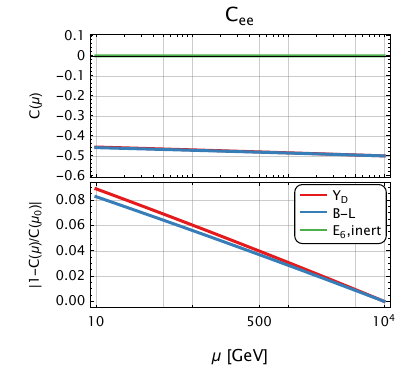}}\hfill
	{\includegraphics[width=\subfigwidth]{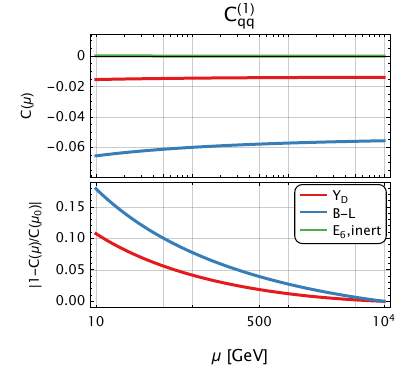}}
	\figcap
\end{figure}
\begin{figure}[p]
  \ContinuedFloat
	\captionsetup[subfigure]{format=plain}
	\captionsetup{list=off,format=cont}
	\centering
	{\includegraphics[width=\subfigwidth]{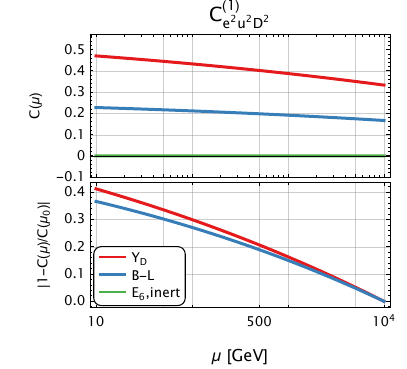}}\hfill
	{\includegraphics[width=\subfigwidth]{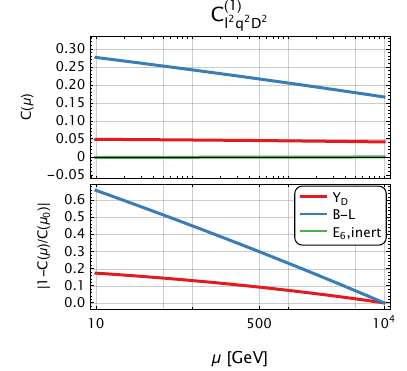}}\\
	{\includegraphics[width=\subfigwidth]{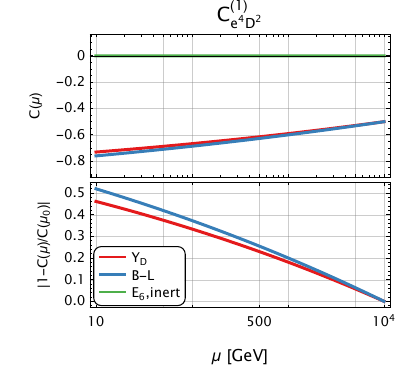}}\hfill
	{\includegraphics[width=\subfigwidth]{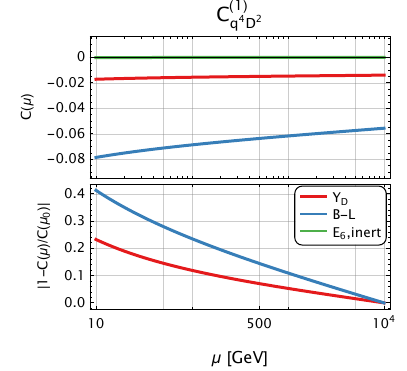}}
	\figcap
	\label{fig:running}
\end{figure}
\endgroup

\subsection{Fit to $Z^{\prime}$ models}

We next study the effect of the RG running on fits to LHC Drell-Yan data using several representative $Z^{\prime}$ models. Each model is defined by a coupling constant $g_D$ and the $Z^{\prime}$ mass which we set to 10 TeV. Two other parameters also appear as discussed earlier in this text and in detail in~\cite{Dawson:2024ozw}: a kinetic mixing with the SM hypercharge gauge group, and a coupling of the $Z^{\prime}$ to the SM Higgs doublet that gives rise to a $Z^{\prime}$ mass contribution. We set both of these coefficients to zero for simplicity in this initial analysis, as discussed in an earlier section. We fit these models to the LHC data at various orders in the $1/\Lambda$ expansion, both with and without the effects of RG running, to illustrate the impact of various terms in the SMEFT expansion on phenomenological results. Assuming a specific $Z^{\prime}$ model at the matching scale generically activates the full set of four-fermion Wilson coefficients.

\begin{figure}[!htbp]
	\centering
	\includegraphics[width=0.47\textwidth]{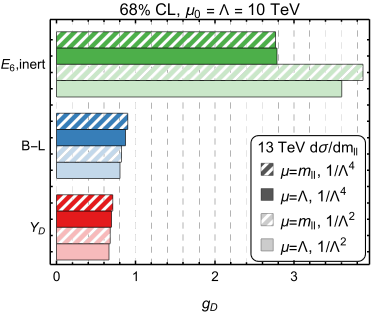}	
	\includegraphics[width=0.47\textwidth]{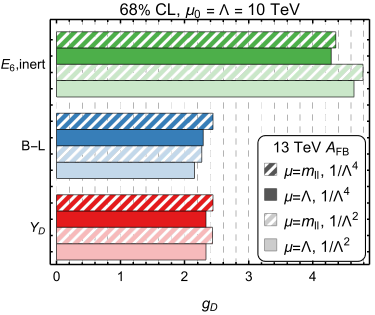}
	\caption{$68\%$ CL fit of $g_D$ with different Z' models using the 13 TeV invariant-mass and $A_{FB}$ data discussed in the main text.
	The initial scale $\mu_0$ is set to the new physics scale $\Lambda=M_{Z'}=10\ \mathrm{TeV}$. The bounds are obtained at different orders in the 
	SMEFT expansion: $1/\Lambda^2$ with and without running, and $1/\Lambda^4$ with and without running. The left panel shows bounds from the invariant mass distribution while the right panel shows the $A_{FB}$ results. }
	\label{fig:1D_fits_models}
\end{figure}

As example datasets we use the 13 TeV invariant mass and forward-backward asymmetry measurements for the Drell-Yan process as reported by the CMS collaboration. The Drell-Yan process is particularly sensitive to semileptonic four-fermion operators which makes this a good choice to illustrate the RG effects obtained here. The details of these datasets are shown in Table~\ref{tab:datasets}. This data was shown to provide strong constraints on semileptonic four-fermion operators in previous work~\cite{Boughezal:2023nhe}, and our study closely follows the details of this analysis. We perform fits to both the $1/\Lambda^2$ and $1/\Lambda^4$ orders in the SMEFT expansion, with and without running. In the fits where the RG running is enabled we choose the renormalization scale of the Wilson coefficients in each $m_{ll}$ bin to be the lower edge of the $m_{ll}$ bin. In Fig.~\ref{fig:1D_fits_models}, we show the 1-d fit of $g_D$ with different Z' models. While the RG running effects can be substantial for individual operators as shown previously, the overall effect on the bound on the coupling parameter $g_D$ that defines each $Z^{\prime}$ model is small. This is due to the fact that the SMEFT deviations are largest in the high-$m_{ll}$ region where the running effects are small.  The largest effects from both $1/\Lambda^4$ effects and RG running are seen for the inert $E_6$ model. The bounds are significantly weaker for this model, primarily due to the limited number of operators activated at the matching scale. We note that the effect of the dimension-8 running is at most a few percent when considering fits to the $A_{FB}$ data, for which the bounds are weaker than the invariant mass distributions. Going from $1/\Lambda^2$ to $1/\Lambda^4$ in the SMEFT expansion has a measurable effect only for the inert $E_6$ model, and almost no effect for the other models. This small impact of the dimension-8 terms within specific $Z^{\prime}$ models was also noted previously in~\cite{Dawson:2024ozw}.

{\setlength{\tabcolsep}{6pt}
\begin{table}[htbp]
	\centering
	\begin{tabular}{c|lccccc}
		\hline
		\hline
		 Experiment & $\sqrt{s}$          & Measurement       & Luminosity               & $m_{ll}^{\rm low}$       & Ref.                 \\ \hline
				 CMS        & $13 \ \mathrm{TeV}$ & $\dv*{\sigma}{m}$ & \makecell{$137 \ \mathrm{fb}^{-1}\ (ee)$ \\$140 \ \mathrm{fb}^{-1}\ (\mu\mu)$} &  \makecell{$200-2210 \ \mathrm{GeV}\ (ee)$\\$210-2290 \ \mathrm{GeV}\ (\mu\mu)$}     & \cite{CMS:2021ctt}     \\
											\hline 
				CMS        & $13 \ {\rm TeV}$ & $A_{\rm FB}$   & $138 \ {\rm fb}^{-1}$ &  $170-1000 \ {\rm GeV}$             & \cite{CMS:2022uul} \\ 											
									\hline		\hline
	\end{tabular}
	\caption{Summary of the Drell-Yan datasets used in this analysis. $m_{ll}^{\rm low}$ denotes the range of the lower edges of the dilepton invariant mass used in this work. More details regarding the treatment of these datasets is given in~\cite{Boughezal:2023nhe}. }
	\label{tab:datasets}
\end{table}}

\section{Conclusions}
\label{sec:conc}

In this manuscript we have derived the renormalization-group equations for dimension-8 four-fermion Wilson coefficients in the SMEFT. These RGEs represent a subset of the higher-order loop corrections needed for precision studies of new physics in the SMEFT framework, and are particularly important when analyzing data from experiments from widely-separated energy scales. We have presented numerical results for the running of both dimension-6 and dimension-8 Wilson coefficients in a selection of UV models containing $Z^{\prime}$ bosons. The corrections from running of the dimension-8 coefficients can reach 50\% or more when running from multi-TeV energies down to the few-GeV scales relevant for lower-energy experiments. To examine the effect of the RG running on fits to LHC data we considered 13 TeV Drell-Yan data in the context of explicit $Z^{\prime}$ models matched to the SMEFT. We found that the running generically has a small impact on the bounds obtained as the fits are dominated by high invariant-mass bins where the effect of the RGEs are minimal. We observe non-negligible effects from dimension-8 RG running only in models where the constraints are weak due to only a small subset of the possible four-fermion operators being activated at the UV matching scale. We consider our work an initial step in the study of the impact of RG evolution at the dimension-8 level in the SMEFT, and we look forward to future work in this direction. In particular the impact of the dimension-8 running when combining LHC measurements with precision measurements at intermediate energies, such as at a future electron-ion collider or circular $e^+e^-$ machine, would be interesting to study. 
{Another important direction for future research is the exploration of two-loop  dimension-6 RG evolution. While the study of two-loop RGEs in SMEFT is still at a very early stage~\cite{Bern:2020ikv,DiNoi:2024ajj,Born:2024mgz}, this line of inquiry could reveal scenarios where the evolution effects of two-loop dimension-6 RGEs become comparable to dimension-8 ones. Initial work in Higgs physics has highlighted potentially impactful effects from these two-loop contributions, underscoring the need for further investigation in this area~\cite{DiNoi:2024ajj}.}

\section*{Acknowledgments}
 R.~B. is supported by the DOE contract DE-AC02-06CH11357.  Y.~H and F.~P. are supported
by the DOE grants DE-FG02-91ER40684 and DE-AC02-06CH11357. This research was supported in part through the computational resources and staff contributions provided for the Quest high performance computing facility at Northwestern University which is jointly supported by the Office of the Provost, the Office for Research, and Northwestern University Information Technology.

\appendix
\counterwithin*{equation}{section}
\renewcommand\theequation{\thesection\arabic{equation}}
\section{RGEs for dimension-8 coefficients assuming MFV}

We present here the RGEs for the dimension-8 four-fermion operators assuming MFV. As discussed in the main text, the evolution contains linear terms coming from dimension-8 coefficients, and quadratic terms coming from the insertion of two dimension-6 operators. {We define the linear dimension-8 contribution to the dimension-8 RGEs as
\begin{align}
	\dot{\bar{C}}=16\pi^2\sum_j\gamma_{ij}^{(8)}C_j^{(8)}
\end{align}
and the quadratic dimension-6 contribution as
\begin{align}
	\dot{\hat{C}}=16\pi^2\sum_{j,k}\gamma_{ijk}^{(6)}C_j^{(6)}C_k^{(6)}. 
\end{align}
The full RGEs are then given by
\begin{align}
	\dot{C}=\dot{\bar{C}}+\dot{\hat{C}}
	\label{eq:fullRGEsplit}
\end{align}}
where $C$ on the left-hand side is a dimension-8 Wilson coefficient. 
{In the following, we present the linear dimension-8 and quadratic dimension-6 contributions separately, organized according to their helicity structure, with $L$
denoting left-handed doublets and $R$
denoting right-handed singlets.}

\subsection{Linear dimension-8 $(\bar LL)(\bar LL)$}
\begin{dgroup}
	\begin{dmath*}
		\dot{\bar{C}}_{l^4D^2}^{(1)}= \frac{11}{12} g_1^2 C_{l^4D^2}^{(1)'}+\frac{5}{12} g_W^2 C_{l^4D^2}^{(1)'}+\frac{13}{12} g_1^2 C_{l^4D^2}^{(2)'}-\frac{5}{12} g_W^2 C_{l^4D^2}^{(2)'}+\frac{61}{12} g_1^2 C_{l^4D^2}^{(1)}+\frac{3}{4} g_W^2 C_{l^4D^2}^{(1)}+\frac{2}{3} g_1^2 N_c C_{l^2d^2D^2}^{(1)}+2 g_1^2 C_{l^2e^2D^2}^{(1)}-\frac{2}{3} g_1^2 N_c C_{l^2q^2D^2}^{(1)}-\frac{4}{3} g_1^2 N_c C_{l^2u^2D^2}^{(1)}-2 g_W^2 N_c C_{l^2q^2D^2}^{(3)}+\frac{21}{4} g_1^2 C_{l^4D^2}^{(2)}-\frac{5}{12} g_W^2 C_{l^4D^2}^{(2)}-g_W^2 N_c C_{l^2q^2D^2}^{(4)}-\frac{1}{3} g_1^2 N_c C_{l^2d^2D^2}^{(2)}-g_1^2 C_{l^2e^2D^2}^{(2)}-\frac{1}{3} g_1^2 N_c C_{l^2q^2D^2}^{(2)}+\frac{2}{3} g_1^2 N_c C_{l^2u^2D^2}^{(2)}
	\end{dmath*}
	\begin{dmath*}
		\dot{\bar{C}}_{l^4D^2}^{(1)'}= \frac{7}{3} g_1^2 C_{l^4D^2}^{(1)'}-\frac{1}{3} g_W^2 C_{l^4D^2}^{(1)'}+2 g_1^2 C_{l^4D^2}^{(2)'}-\frac{1}{3} g_W^2 C_{l^4D^2}^{(2)'}+\frac{1}{2} g_W^2 C_{l^4D^2}^{(1)}+g_W^2 N_c C_{l^2q^2D^2}^{(3)}-\frac{7}{2} g_W^2 C_{l^4D^2}^{(2)}+\frac{1}{2} g_W^2 N_c C_{l^2q^2D^2}^{(4)}
	\end{dmath*}
	\begin{dmath*}
		\dot{\bar{C}}_{l^4D^2}^{(2)}= \frac{4}{3} g_1^2 C_{l^4D^2}^{(1)}+\frac{5}{2} g_W^2 C_{l^4D^2}^{(1)}+3 g_1^2 C_{l^4D^2}^{(2)}+\frac{35}{6} g_W^2 C_{l^4D^2}^{(2)}
	\end{dmath*}
	\begin{dmath*}
		\dot{\bar{C}}_{l^4D^2}^{(2)'}= \frac{4}{3} g_1^2 C_{l^4D^2}^{(1)'}+\frac{7}{3} g_W^2 C_{l^4D^2}^{(1)'}+3 g_1^2 C_{l^4D^2}^{(2)'}+7 g_W^2 C_{l^4D^2}^{(2)'}-\frac{1}{2} g_W^2 C_{l^4D^2}^{(1)}-g_W^2 N_c C_{l^2q^2D^2}^{(3)}+\frac{7}{2} g_W^2 C_{l^4D^2}^{(2)}-\frac{1}{2} g_W^2 N_c C_{l^2q^2D^2}^{(4)}
	\end{dmath*}
	\begin{dmath*}
		\dot{\bar{C}}_{q^4D^2}^{(1)}= \frac{4}{27} g_1^2 N_c C_{q^4D^2}^{(1)'}-\frac{1}{18} g_1^2 C_{q^4D^2}^{(1)'}+\frac{g_s^2 C_{q^4D^2}^{(1)'}}{N_c}+\frac{5}{6} g_s^2 C_{q^4D^2}^{(1)'}-\frac{1}{6} g_1^2 C_{q^4D^2}^{(3)'}+\frac{3 g_s^2 C_{q^4D^2}^{(3)'}}{N_c}+\frac{5}{2} g_s^2 C_{q^4D^2}^{(3)'}+\frac{2}{27} g_1^2 N_c C_{q^4D^2}^{(2)'}+\frac{7}{18} g_1^2 C_{q^4D^2}^{(2)'}+\frac{7}{6} g_1^2 C_{q^4D^2}^{(4)'}-\frac{7 g_s^2 C_{q^4D^2}^{(2)'}}{N_c}-\frac{21 g_s^2 C_{q^4D^2}^{(4)'}}{N_c}+\frac{11}{6} g_s^2 C_{q^4D^2}^{(2)'}+\frac{11}{2} g_s^2 C_{q^4D^2}^{(4)'}+\frac{4}{9} g_1^2 N_c C_{q^4D^2}^{(1)}+\frac{13}{54} g_1^2 C_{q^4D^2}^{(1)}-\frac{13 g_s^2 C_{q^4D^2}^{(1)}}{3 N_c}-\frac{2}{9} g_1^2 N_c C_{q^2d^2D^2}^{(1)}-\frac{2}{3} g_1^2 C_{q^2e^2D^2}^{(1)}-\frac{2}{3} g_1^2 C_{l^2q^2D^2}^{(1)}+\frac{4}{9} g_1^2 N_c C_{q^2u^2D^2}^{(1)}-\frac{1}{18} g_1^2 C_{q^4D^2}^{(3)}+\frac{g_s^2 C_{q^4D^2}^{(3)}}{N_c}+7 g_W^2 C_{q^4D^2}^{(3)}-\frac{g_s^2 C_{q^2d^2D^2}^{(3)}}{N_c}-\frac{g_s^2 C_{q^2u^2D^2}^{(3)}}{N_c}+\frac{g_s^2 C_{q^2d^2D^2}^{(4)}}{2 N_c}+\frac{2}{9} g_1^2 N_c C_{q^4D^2}^{(2)}+\frac{19}{54} g_1^2 C_{q^4D^2}^{(2)}+\frac{7}{18} g_1^2 C_{q^4D^2}^{(4)}-\frac{8}{3} g_s^2 N_c C_{q^4D^2}^{(2)}-\frac{19 g_s^2 C_{q^4D^2}^{(2)}}{3 N_c}-\frac{7 g_s^2 C_{q^4D^2}^{(4)}}{N_c}+\frac{g_s^2 C_{q^2u^2D^2}^{(4)}}{2 N_c}-4 g_W^2 C_{q^4D^2}^{(2)}+10 g_W^2 C_{q^4D^2}^{(4)}+\frac{1}{9} g_1^2 N_c C_{q^2d^2D^2}^{(2)}+\frac{1}{3} g_1^2 C_{q^2e^2D^2}^{(2)}-\frac{1}{3} g_1^2 C_{l^2q^2D^2}^{(2)}-\frac{2}{9} g_1^2 N_c C_{q^2u^2D^2}^{(2)}
	\end{dmath*}
	\begin{dmath*}
		\dot{\bar{C}}_{q^4D^2}^{(1)'}= \frac{7}{27} g_1^2 C_{q^4D^2}^{(1)'}-\frac{14 g_s^2 C_{q^4D^2}^{(1)'}}{3 N_c}+\frac{1}{4} g_s^2 C_{q^4D^2}^{(1)'}+\frac{3}{4} g_s^2 C_{q^4D^2}^{(3)'}+7 g_W^2 C_{q^4D^2}^{(3)'}+\frac{2}{9} g_1^2 C_{q^4D^2}^{(2)'}-\frac{8}{3} g_s^2 N_c C_{q^4D^2}^{(2)'}-\frac{4 g_s^2 C_{q^4D^2}^{(2)'}}{N_c}-\frac{7}{4} g_s^2 C_{q^4D^2}^{(2)'}-\frac{21}{4} g_s^2 C_{q^4D^2}^{(4)'}-4 g_W^2 C_{q^4D^2}^{(2)'}+10 g_W^2 C_{q^4D^2}^{(4)'}+\frac{11}{12} g_s^2 C_{q^4D^2}^{(1)}+\frac{11}{4} g_s^2 C_{q^4D^2}^{(3)}-\frac{1}{4} g_s^2 C_{q^2d^2D^2}^{(3)}-\frac{1}{4} g_s^2 C_{q^2u^2D^2}^{(3)}+\frac{1}{8} g_s^2 C_{q^2d^2D^2}^{(4)}+\frac{5}{4} g_s^2 C_{q^4D^2}^{(2)}+\frac{15}{4} g_s^2 C_{q^4D^2}^{(4)}+\frac{1}{8} g_s^2 C_{q^2u^2D^2}^{(4)}
	\end{dmath*}
	\begin{dmath*}
		\dot{\bar{C}}_{q^4D^2}^{(2)}= \frac{11}{6} g_s^2 C_{q^4D^2}^{(1)'}+\frac{11}{2} g_s^2 C_{q^4D^2}^{(3)'}+\frac{17}{6} g_s^2 C_{q^4D^2}^{(2)'}+\frac{17}{2} g_s^2 C_{q^4D^2}^{(4)'}+\frac{4}{27} g_1^2 C_{q^4D^2}^{(1)}-\frac{8 g_s^2 C_{q^4D^2}^{(1)}}{3 N_c}+4 g_W^2 C_{q^4D^2}^{(3)}+\frac{1}{3} g_1^2 C_{q^4D^2}^{(2)}+\frac{4}{3} g_s^2 N_c C_{q^4D^2}^{(2)}-\frac{6 g_s^2 C_{q^4D^2}^{(2)}}{N_c}+2 g_W^2 C_{q^4D^2}^{(2)}+7 g_W^2 C_{q^4D^2}^{(4)}
	\end{dmath*}
	\begin{dmath*}
		\dot{\bar{C}}_{q^4D^2}^{(2)'}= \frac{4}{27} g_1^2 C_{q^4D^2}^{(1)'}-\frac{8 g_s^2 C_{q^4D^2}^{(1)'}}{3 N_c}-\frac{1}{4} g_s^2 C_{q^4D^2}^{(1)'}-\frac{3}{4} g_s^2 C_{q^4D^2}^{(3)'}+4 g_W^2 C_{q^4D^2}^{(3)'}+\frac{1}{3} g_1^2 C_{q^4D^2}^{(2)'}+\frac{4}{3} g_s^2 N_c C_{q^4D^2}^{(2)'}-\frac{6 g_s^2 C_{q^4D^2}^{(2)'}}{N_c}+\frac{7}{4} g_s^2 C_{q^4D^2}^{(2)'}+\frac{21}{4} g_s^2 C_{q^4D^2}^{(4)'}+2 g_W^2 C_{q^4D^2}^{(2)'}+7 g_W^2 C_{q^4D^2}^{(4)'}+\frac{7}{4} g_s^2 C_{q^4D^2}^{(1)}+\frac{21}{4} g_s^2 C_{q^4D^2}^{(3)}+\frac{1}{4} g_s^2 C_{q^2d^2D^2}^{(3)}+\frac{1}{4} g_s^2 C_{q^2u^2D^2}^{(3)}-\frac{1}{8} g_s^2 C_{q^2d^2D^2}^{(4)}+\frac{41}{12} g_s^2 C_{q^4D^2}^{(2)}+\frac{41}{4} g_s^2 C_{q^4D^2}^{(4)}-\frac{1}{8} g_s^2 C_{q^2u^2D^2}^{(4)}
	\end{dmath*}
	\begin{dmath*}
		\dot{\bar{C}}_{q^4D^2}^{(3)}= \frac{5}{6} g_s^2 C_{q^4D^2}^{(1)'}-\frac{1}{2} g_W^2 C_{q^4D^2}^{(1)'}-\frac{5}{6} g_s^2 C_{q^4D^2}^{(3)'}+\frac{4}{3} g_W^2 N_c C_{q^4D^2}^{(3)'}+\frac{1}{2} g_W^2 C_{q^4D^2}^{(3)'}+\frac{11}{6} g_s^2 C_{q^4D^2}^{(2)'}-\frac{11}{6} g_s^2 C_{q^4D^2}^{(4)'}+\frac{2}{3} g_W^2 N_c C_{q^4D^2}^{(4)'}+\frac{7}{2} g_W^2 C_{q^4D^2}^{(2)'}-\frac{7}{2} g_W^2 C_{q^4D^2}^{(4)'}+\frac{13}{6} g_W^2 C_{q^4D^2}^{(1)}+\frac{7}{27} g_1^2 C_{q^4D^2}^{(3)}-\frac{14 g_s^2 C_{q^4D^2}^{(3)}}{3 N_c}+4 g_W^2 N_c C_{q^4D^2}^{(3)}-\frac{19}{6} g_W^2 C_{q^4D^2}^{(3)}+2 g_W^2 C_{l^2q^2D^2}^{(3)}+\frac{2}{9} g_1^2 C_{q^4D^2}^{(4)}-\frac{8}{3} g_s^2 N_c C_{q^4D^2}^{(4)}-\frac{4 g_s^2 C_{q^4D^2}^{(4)}}{N_c}+g_W^2 C_{l^2q^2D^2}^{(4)}+2 g_W^2 N_c C_{q^4D^2}^{(4)}+\frac{9}{2} g_W^2 C_{q^4D^2}^{(2)}-\frac{7}{6} g_W^2 C_{q^4D^2}^{(4)}
	\end{dmath*}
	\begin{dmath*}
		\dot{\bar{C}}_{q^4D^2}^{(3)'}= \frac{1}{4} g_s^2 C_{q^4D^2}^{(1)'}+\frac{7}{3} g_W^2 C_{q^4D^2}^{(1)'}+\frac{7}{27} g_1^2 C_{q^4D^2}^{(3)'}-\frac{14 g_s^2 C_{q^4D^2}^{(3)'}}{3 N_c}+\frac{3}{4} g_s^2 C_{q^4D^2}^{(3)'}-\frac{10}{3} g_W^2 C_{q^4D^2}^{(3)'}+\frac{2}{9} g_1^2 C_{q^4D^2}^{(4)'}-\frac{8}{3} g_s^2 N_c C_{q^4D^2}^{(4)'}-\frac{4 g_s^2 C_{q^4D^2}^{(4)'}}{N_c}-\frac{7}{4} g_s^2 C_{q^4D^2}^{(2)'}-\frac{21}{4} g_s^2 C_{q^4D^2}^{(4)'}+\frac{10}{3} g_W^2 C_{q^4D^2}^{(2)'}+\frac{11}{12} g_s^2 C_{q^4D^2}^{(1)}-\frac{7}{12} g_s^2 C_{q^4D^2}^{(3)}-\frac{1}{4} g_s^2 C_{q^2d^2D^2}^{(3)}-\frac{1}{4} g_s^2 C_{q^2u^2D^2}^{(3)}+\frac{1}{8} g_s^2 C_{q^2d^2D^2}^{(4)}+\frac{5}{4} g_s^2 C_{q^4D^2}^{(2)}-\frac{43}{12} g_s^2 C_{q^4D^2}^{(4)}+\frac{1}{8} g_s^2 C_{q^2u^2D^2}^{(4)}
	\end{dmath*}
	\begin{dmath*}
		\dot{\bar{C}}_{q^4D^2}^{(4)}= \frac{11}{6} g_s^2 C_{q^4D^2}^{(1)'}-\frac{11}{6} g_s^2 C_{q^4D^2}^{(3)'}+\frac{17}{6} g_s^2 C_{q^4D^2}^{(2)'}-\frac{17}{6} g_s^2 C_{q^4D^2}^{(4)'}+\frac{4}{3} g_W^2 C_{q^4D^2}^{(1)}+\frac{4}{27} g_1^2 C_{q^4D^2}^{(3)}-\frac{8 g_s^2 C_{q^4D^2}^{(3)}}{3 N_c}-\frac{4}{3} g_W^2 C_{q^4D^2}^{(3)}+\frac{1}{3} g_1^2 C_{q^4D^2}^{(4)}+\frac{4}{3} g_s^2 N_c C_{q^4D^2}^{(4)}-\frac{6 g_s^2 C_{q^4D^2}^{(4)}}{N_c}+\frac{7}{3} g_W^2 C_{q^4D^2}^{(2)}-8 g_W^2 C_{q^4D^2}^{(4)}
	\end{dmath*}
	\begin{dmath*}
		\dot{\bar{C}}_{q^4D^2}^{(4)'}= -\frac{1}{4} g_s^2 C_{q^4D^2}^{(1)'}+\frac{4}{3} g_W^2 C_{q^4D^2}^{(1)'}+\frac{4}{27} g_1^2 C_{q^4D^2}^{(3)'}-\frac{8 g_s^2 C_{q^4D^2}^{(3)'}}{3 N_c}-\frac{3}{4} g_s^2 C_{q^4D^2}^{(3)'}-\frac{4}{3} g_W^2 C_{q^4D^2}^{(3)'}+\frac{1}{3} g_1^2 C_{q^4D^2}^{(4)'}+\frac{4}{3} g_s^2 N_c C_{q^4D^2}^{(4)'}-\frac{6 g_s^2 C_{q^4D^2}^{(4)'}}{N_c}+\frac{7}{4} g_s^2 C_{q^4D^2}^{(2)'}+\frac{21}{4} g_s^2 C_{q^4D^2}^{(4)'}+\frac{7}{3} g_W^2 C_{q^4D^2}^{(2)'}-8 g_W^2 C_{q^4D^2}^{(4)'}+\frac{7}{4} g_s^2 C_{q^4D^2}^{(1)}-\frac{25}{12} g_s^2 C_{q^4D^2}^{(3)}+\frac{1}{4} g_s^2 C_{q^2d^2D^2}^{(3)}+\frac{1}{4} g_s^2 C_{q^2u^2D^2}^{(3)}-\frac{1}{8} g_s^2 C_{q^2d^2D^2}^{(4)}+\frac{41}{12} g_s^2 C_{q^4D^2}^{(2)}-\frac{13}{12} g_s^2 C_{q^4D^2}^{(4)}-\frac{1}{8} g_s^2 C_{q^2u^2D^2}^{(4)}
	\end{dmath*}
	\begin{dmath*}
		\dot{\bar{C}}_{l^2q^2D^2}^{(1)}= -\frac{7}{18} g_1^2 C_{l^4D^2}^{(1)'}-\frac{4}{9} g_1^2 N_c C_{q^4D^2}^{(1)'}+\frac{1}{6} g_1^2 C_{q^4D^2}^{(1)'}+\frac{1}{2} g_1^2 C_{q^4D^2}^{(3)'}-\frac{11}{18} g_1^2 C_{l^4D^2}^{(2)'}-\frac{2}{9} g_1^2 N_c C_{q^4D^2}^{(2)'}-\frac{7}{6} g_1^2 C_{q^4D^2}^{(2)'}-\frac{7}{2} g_1^2 C_{q^4D^2}^{(4)'}-\frac{7}{6} g_1^2 C_{l^4D^2}^{(1)}-\frac{4}{3} g_1^2 N_c C_{q^4D^2}^{(1)}+\frac{1}{18} g_1^2 C_{q^4D^2}^{(1)}-\frac{2}{9} g_1^2 N_c C_{l^2d^2D^2}^{(1)}+\frac{2}{3} g_1^2 N_c C_{q^2d^2D^2}^{(1)}-\frac{2}{3} g_1^2 C_{l^2e^2D^2}^{(1)}+2 g_1^2 C_{q^2e^2D^2}^{(1)}+\frac{2}{9} g_1^2 N_c C_{l^2q^2D^2}^{(1)}+\frac{4}{9} g_1^2 N_c C_{l^2u^2D^2}^{(1)}+\frac{11}{9} g_1^2 C_{l^2q^2D^2}^{(1)}-\frac{4}{3} g_1^2 N_c C_{q^2u^2D^2}^{(1)}+\frac{1}{6} g_1^2 C_{q^4D^2}^{(3)}-\frac{11}{6} g_1^2 C_{l^4D^2}^{(2)}-\frac{2}{3} g_1^2 N_c C_{q^4D^2}^{(2)}-\frac{7}{18} g_1^2 C_{q^4D^2}^{(2)}-\frac{7}{6} g_1^2 C_{q^4D^2}^{(4)}+\frac{1}{9} g_1^2 N_c C_{l^2d^2D^2}^{(2)}-\frac{1}{3} g_1^2 N_c C_{q^2d^2D^2}^{(2)}+\frac{1}{3} g_1^2 C_{l^2e^2D^2}^{(2)}-g_1^2 C_{q^2e^2D^2}^{(2)}+\frac{1}{9} g_1^2 N_c C_{l^2q^2D^2}^{(2)}-\frac{2}{9} g_1^2 N_c C_{l^2u^2D^2}^{(2)}-\frac{23}{27} g_1^2 C_{l^2q^2D^2}^{(2)}+\frac{2}{3} g_1^2 N_c C_{q^2u^2D^2}^{(2)}-\frac{4}{3} g_s^2 N_c C_{l^2q^2D^2}^{(2)}+\frac{4 g_s^2 C_{l^2q^2D^2}^{(2)}}{3 N_c}-4 g_W^2 C_{l^2q^2D^2}^{(2)}
	\end{dmath*}
	\begin{dmath*}
		\dot{\bar{C}}_{l^2q^2D^2}^{(2)}= -\frac{4}{9} g_1^2 C_{l^2q^2D^2}^{(1)}-\frac{11}{27} g_1^2 C_{l^2q^2D^2}^{(2)}+\frac{2}{3} g_s^2 N_c C_{l^2q^2D^2}^{(2)}-\frac{2 g_s^2 C_{l^2q^2D^2}^{(2)}}{3 N_c}+2 g_W^2 C_{l^2q^2D^2}^{(2)}
	\end{dmath*}
	\begin{dmath*}
		\dot{\bar{C}}_{l^2q^2D^2}^{(3)}= -\frac{1}{6} g_W^2 C_{l^4D^2}^{(1)'}-\frac{1}{2} g_W^2 C_{q^4D^2}^{(1)'}+\frac{4}{3} g_W^2 N_c C_{q^4D^2}^{(3)'}+\frac{1}{2} g_W^2 C_{q^4D^2}^{(3)'}+\frac{7}{6} g_W^2 C_{l^4D^2}^{(2)'}+\frac{2}{3} g_W^2 N_c C_{q^4D^2}^{(4)'}+\frac{7}{2} g_W^2 C_{q^4D^2}^{(2)'}-\frac{7}{2} g_W^2 C_{q^4D^2}^{(4)'}-\frac{1}{2} g_W^2 C_{l^4D^2}^{(1)}-\frac{1}{6} g_W^2 C_{q^4D^2}^{(1)}+\frac{7}{3} g_W^2 C_{l^2q^2D^2}^{(1)}+4 g_W^2 N_c C_{q^4D^2}^{(3)}+\frac{1}{6} g_W^2 C_{q^4D^2}^{(3)}-\frac{7}{9} g_1^2 C_{l^2q^2D^2}^{(3)}+2 g_W^2 N_c C_{l^2q^2D^2}^{(3)}+\frac{29}{3} g_W^2 C_{l^2q^2D^2}^{(3)}-\frac{50}{27} g_1^2 C_{l^2q^2D^2}^{(4)}-\frac{4}{3} g_s^2 N_c C_{l^2q^2D^2}^{(4)}+\frac{4 g_s^2 C_{l^2q^2D^2}^{(4)}}{3 N_c}+\frac{7}{2} g_W^2 C_{l^4D^2}^{(2)}+g_W^2 N_c C_{l^2q^2D^2}^{(4)}+3 g_W^2 C_{l^2q^2D^2}^{(4)}+2 g_W^2 N_c C_{q^4D^2}^{(4)}+\frac{7}{6} g_W^2 C_{q^4D^2}^{(2)}-\frac{7}{6} g_W^2 C_{q^4D^2}^{(4)}+\frac{10}{3} g_W^2 C_{l^2q^2D^2}^{(2)}
	\end{dmath*}
	\begin{dmath*}
		\dot{\bar{C}}_{l^2q^2D^2}^{(4)}= \frac{4}{3} g_W^2 C_{l^2q^2D^2}^{(1)}-\frac{4}{9} g_1^2 C_{l^2q^2D^2}^{(3)}+\frac{2}{3} g_W^2 C_{l^2q^2D^2}^{(3)}-\frac{11}{27} g_1^2 C_{l^2q^2D^2}^{(4)}+\frac{2}{3} g_s^2 N_c C_{l^2q^2D^2}^{(4)}-\frac{2 g_s^2 C_{l^2q^2D^2}^{(4)}}{3 N_c}+9 g_W^2 C_{l^2q^2D^2}^{(4)}+\frac{7}{3} g_W^2 C_{l^2q^2D^2}^{(2)}
	\end{dmath*}
\end{dgroup}
\subsection{Linear dimension-8 $(\bar LL)(\bar RR)$}
\begin{dgroup}
	\begin{dmath*}
		\dot{\bar{C}}_{l^2d^2D^2}^{(1)}= \frac{4}{9} g_1^2 N_c C_{d^4D^2}^{(1)'}-\frac{1}{3} g_1^2 C_{d^4D^2}^{(1)'}+\frac{7}{9} g_1^2 C_{l^4D^2}^{(1)'}+\frac{2}{9} g_1^2 N_c C_{d^4D^2}^{(2)'}+\frac{7}{3} g_1^2 C_{d^4D^2}^{(2)'}+\frac{11}{9} g_1^2 C_{l^4D^2}^{(2)'}+\frac{4}{3} g_1^2 N_c C_{d^4D^2}^{(1)}-\frac{1}{9} g_1^2 C_{d^4D^2}^{(1)}+\frac{7}{3} g_1^2 C_{l^4D^2}^{(1)}+2 g_1^2 C_{e^2d^2D^2}^{(1)}+\frac{4}{9} g_1^2 N_c C_{l^2d^2D^2}^{(1)}+\frac{4}{9} g_1^2 C_{l^2d^2D^2}^{(1)}-\frac{2}{3} g_1^2 N_c C_{q^2d^2D^2}^{(1)}-\frac{4}{3} g_1^2 N_c C_{u^2d^2D^2}^{(1)}+\frac{4}{3} g_1^2 C_{l^2e^2D^2}^{(1)}-\frac{4}{9} g_1^2 N_c C_{l^2q^2D^2}^{(1)}-\frac{8}{9} g_1^2 N_c C_{l^2u^2D^2}^{(1)}+\frac{2}{3} g_1^2 N_c C_{d^4D^2}^{(2)}+\frac{7}{9} g_1^2 C_{d^4D^2}^{(2)}+\frac{11}{3} g_1^2 C_{l^4D^2}^{(2)}+g_1^2 C_{e^2d^2D^2}^{(2)}-\frac{2}{9} g_1^2 N_c C_{l^2d^2D^2}^{(2)}+\frac{59}{27} g_1^2 C_{l^2d^2D^2}^{(2)}+\frac{1}{3} g_1^2 N_c C_{q^2d^2D^2}^{(2)}-\frac{2}{3} g_1^2 N_c C_{u^2d^2D^2}^{(2)}+\frac{4}{3} g_s^2 N_c C_{l^2d^2D^2}^{(2)}-\frac{4 g_s^2 C_{l^2d^2D^2}^{(2)}}{3 N_c}+2 g_W^2 C_{l^2d^2D^2}^{(2)}-\frac{2}{3} g_1^2 C_{l^2e^2D^2}^{(2)}-\frac{2}{9} g_1^2 N_c C_{l^2q^2D^2}^{(2)}+\frac{4}{9} g_1^2 N_c C_{l^2u^2D^2}^{(2)}
	\end{dmath*}
	\begin{dmath*}
		\dot{\bar{C}}_{l^2d^2D^2}^{(2)}= \frac{8}{9} g_1^2 C_{l^2d^2D^2}^{(1)}-\frac{29}{27} g_1^2 C_{l^2d^2D^2}^{(2)}+\frac{2}{3} g_s^2 N_c C_{l^2d^2D^2}^{(2)}-\frac{2 g_s^2 C_{l^2d^2D^2}^{(2)}}{3 N_c}+g_W^2 C_{l^2d^2D^2}^{(2)}
	\end{dmath*}
	\begin{dmath*}
		\dot{\bar{C}}_{l^2e^2D^2}^{(1)}= \frac{4}{3} g_1^2 C_{l^4D^2}^{(1)'}+\frac{2}{3} g_1^2 C_{l^4D^2}^{(2)'}+3 g_1^2 C_{e^4D^2}^{(1)}+4 g_1^2 C_{l^4D^2}^{(1)}+\frac{2}{3} g_1^2 N_c C_{e^2d^2D^2}^{(1)}+\frac{4}{3} g_1^2 N_c C_{l^2d^2D^2}^{(1)}+\frac{4}{3} g_1^2 C_{l^2e^2D^2}^{(1)}-\frac{2}{3} g_1^2 N_c C_{q^2e^2D^2}^{(1)}-\frac{4}{3} g_1^2 N_c C_{e^2u^2D^2}^{(1)}-\frac{4}{3} g_1^2 N_c C_{l^2q^2D^2}^{(1)}-\frac{8}{3} g_1^2 N_c C_{l^2u^2D^2}^{(1)}+2 g_1^2 C_{l^4D^2}^{(2)}+\frac{1}{3} g_1^2 N_c C_{e^2d^2D^2}^{(2)}-\frac{2}{3} g_1^2 N_c C_{l^2d^2D^2}^{(2)}+7 g_1^2 C_{l^2e^2D^2}^{(2)}+\frac{1}{3} g_1^2 N_c C_{q^2e^2D^2}^{(2)}-\frac{2}{3} g_1^2 N_c C_{e^2u^2D^2}^{(2)}+2 g_W^2 C_{l^2e^2D^2}^{(2)}-\frac{2}{3} g_1^2 N_c C_{l^2q^2D^2}^{(2)}+\frac{4}{3} g_1^2 N_c C_{l^2u^2D^2}^{(2)}
	\end{dmath*}
	\begin{dmath*}
		\dot{\bar{C}}_{l^2e^2D^2}^{(2)}= \frac{8}{3} g_1^2 C_{l^2e^2D^2}^{(1)}-3 g_1^2 C_{l^2e^2D^2}^{(2)}+g_W^2 C_{l^2e^2D^2}^{(2)}
	\end{dmath*}
	\begin{dmath*}
		\dot{\bar{C}}_{l^2u^2D^2}^{(1)}= -\frac{14}{9} g_1^2 C_{l^4D^2}^{(1)'}-\frac{8}{9} g_1^2 N_c C_{u^4D^2}^{(1)'}+\frac{2}{3} g_1^2 C_{u^4D^2}^{(1)'}-\frac{22}{9} g_1^2 C_{l^4D^2}^{(2)'}-\frac{4}{9} g_1^2 N_c C_{u^4D^2}^{(2)'}-\frac{14}{3} g_1^2 C_{u^4D^2}^{(2)'}-\frac{14}{3} g_1^2 C_{l^4D^2}^{(1)}-\frac{8}{3} g_1^2 N_c C_{u^4D^2}^{(1)}+\frac{2}{9} g_1^2 C_{u^4D^2}^{(1)}-\frac{8}{9} g_1^2 N_c C_{l^2d^2D^2}^{(1)}+\frac{2}{3} g_1^2 N_c C_{u^2d^2D^2}^{(1)}-\frac{8}{3} g_1^2 C_{l^2e^2D^2}^{(1)}+2 g_1^2 C_{e^2u^2D^2}^{(1)}+\frac{8}{9} g_1^2 N_c C_{l^2q^2D^2}^{(1)}+\frac{16}{9} g_1^2 N_c C_{l^2u^2D^2}^{(1)}+\frac{46}{9} g_1^2 C_{l^2u^2D^2}^{(1)}-\frac{2}{3} g_1^2 N_c C_{q^2u^2D^2}^{(1)}-\frac{22}{3} g_1^2 C_{l^4D^2}^{(2)}-\frac{4}{3} g_1^2 N_c C_{u^4D^2}^{(2)}-\frac{14}{9} g_1^2 C_{u^4D^2}^{(2)}+\frac{4}{9} g_1^2 N_c C_{l^2d^2D^2}^{(2)}+\frac{1}{3} g_1^2 N_c C_{u^2d^2D^2}^{(2)}+\frac{4}{3} g_1^2 C_{l^2e^2D^2}^{(2)}+g_1^2 C_{e^2u^2D^2}^{(2)}+\frac{4}{9} g_1^2 N_c C_{l^2q^2D^2}^{(2)}-\frac{8}{9} g_1^2 N_c C_{l^2u^2D^2}^{(2)}-\frac{97}{27} g_1^2 C_{l^2u^2D^2}^{(2)}+\frac{1}{3} g_1^2 N_c C_{q^2u^2D^2}^{(2)}+\frac{4}{3} g_s^2 N_c C_{l^2u^2D^2}^{(2)}-\frac{4 g_s^2 C_{l^2u^2D^2}^{(2)}}{3 N_c}+2 g_W^2 C_{l^2u^2D^2}^{(2)}
	\end{dmath*}
	\begin{dmath*}
		\dot{\bar{C}}_{l^2u^2D^2}^{(2)}= -\frac{16}{9} g_1^2 C_{l^2u^2D^2}^{(1)}+\frac{109}{27} g_1^2 C_{l^2u^2D^2}^{(2)}+\frac{2}{3} g_s^2 N_c C_{l^2u^2D^2}^{(2)}-\frac{2 g_s^2 C_{l^2u^2D^2}^{(2)}}{3 N_c}+g_W^2 C_{l^2u^2D^2}^{(2)}
	\end{dmath*}
	\begin{dmath*}
		\dot{\bar{C}}_{q^2d^2D^2}^{(1)}= -\frac{4}{27} g_1^2 N_c C_{d^4D^2}^{(1)'}+\frac{1}{9} g_1^2 C_{d^4D^2}^{(1)'}-\frac{8}{27} g_1^2 N_c C_{q^4D^2}^{(1)'}+\frac{1}{9} g_1^2 C_{q^4D^2}^{(1)'}+\frac{1}{3} g_1^2 C_{q^4D^2}^{(3)'}-\frac{2}{27} g_1^2 N_c C_{d^4D^2}^{(2)'}-\frac{7}{9} g_1^2 C_{d^4D^2}^{(2)'}-\frac{4}{27} g_1^2 N_c C_{q^4D^2}^{(2)'}-\frac{7}{9} g_1^2 C_{q^4D^2}^{(2)'}-\frac{7}{3} g_1^2 C_{q^4D^2}^{(4)'}-\frac{4}{9} g_1^2 N_c C_{d^4D^2}^{(1)}+\frac{1}{27} g_1^2 C_{d^4D^2}^{(1)}-\frac{8}{9} g_1^2 N_c C_{q^4D^2}^{(1)}+\frac{1}{27} g_1^2 C_{q^4D^2}^{(1)}-\frac{2}{3} g_1^2 C_{e^2d^2D^2}^{(1)}-\frac{2}{3} g_1^2 C_{l^2d^2D^2}^{(1)}+\frac{2}{3} g_1^2 N_c C_{q^2d^2D^2}^{(1)}+\frac{4}{9} g_1^2 N_c C_{u^2d^2D^2}^{(1)}+\frac{14}{27} g_1^2 C_{q^2d^2D^2}^{(1)}+\frac{4}{3} g_1^2 C_{q^2e^2D^2}^{(1)}+\frac{4}{3} g_1^2 C_{l^2q^2D^2}^{(1)}-\frac{8}{9} g_1^2 N_c C_{q^2u^2D^2}^{(1)}+\frac{1}{9} g_1^2 C_{q^4D^2}^{(3)}+\frac{7 g_s^2 C_{q^2d^2D^2}^{(3)}}{3 N_c^2}-\frac{7}{3} g_s^2 C_{q^2d^2D^2}^{(3)}-\frac{2}{9} g_1^2 N_c C_{d^4D^2}^{(2)}-\frac{7}{27} g_1^2 C_{d^4D^2}^{(2)}-\frac{10 g_s^2 C_{q^2d^2D^2}^{(4)}}{3 N_c^2}+\frac{10}{3} g_s^2 C_{q^2d^2D^2}^{(4)}-\frac{4}{9} g_1^2 N_c C_{q^4D^2}^{(2)}-\frac{7}{27} g_1^2 C_{q^4D^2}^{(2)}-\frac{7}{9} g_1^2 C_{q^4D^2}^{(4)}-\frac{1}{3} g_1^2 C_{e^2d^2D^2}^{(2)}+\frac{1}{3} g_1^2 C_{l^2d^2D^2}^{(2)}-\frac{1}{3} g_1^2 N_c C_{q^2d^2D^2}^{(2)}+\frac{2}{9} g_1^2 N_c C_{u^2d^2D^2}^{(2)}-\frac{10}{27} g_1^2 C_{q^2d^2D^2}^{(2)}+\frac{8}{3} g_s^2 N_c C_{q^2d^2D^2}^{(2)}-\frac{8 g_s^2 C_{q^2d^2D^2}^{(2)}}{3 N_c}+2 g_W^2 C_{q^2d^2D^2}^{(2)}-\frac{2}{3} g_1^2 C_{q^2e^2D^2}^{(2)}+\frac{2}{3} g_1^2 C_{l^2q^2D^2}^{(2)}+\frac{4}{9} g_1^2 N_c C_{q^2u^2D^2}^{(2)}
	\end{dmath*}
	\begin{dmath*}
		\dot{\bar{C}}_{q^2d^2D^2}^{(2)}= -\frac{8}{27} g_1^2 C_{q^2d^2D^2}^{(1)}-\frac{4 g_s^2 C_{q^2d^2D^2}^{(3)}}{3 N_c^2}+\frac{4}{3} g_s^2 C_{q^2d^2D^2}^{(3)}+\frac{7 g_s^2 C_{q^2d^2D^2}^{(4)}}{3 N_c^2}-\frac{7}{3} g_s^2 C_{q^2d^2D^2}^{(4)}+\frac{19}{27} g_1^2 C_{q^2d^2D^2}^{(2)}+\frac{4}{3} g_s^2 N_c C_{q^2d^2D^2}^{(2)}-\frac{4 g_s^2 C_{q^2d^2D^2}^{(2)}}{3 N_c}+g_W^2 C_{q^2d^2D^2}^{(2)}
	\end{dmath*}
	\begin{dmath*}
		\dot{\bar{C}}_{q^2d^2D^2}^{(3)}= -2 g_s^2 C_{d^4D^2}^{(1)'}-2 g_s^2 C_{q^4D^2}^{(1)'}-6 g_s^2 C_{q^4D^2}^{(3)'}+14 g_s^2 C_{d^4D^2}^{(2)'}+14 g_s^2 C_{q^4D^2}^{(2)'}+42 g_s^2 C_{q^4D^2}^{(4)'}-\frac{2}{3} g_s^2 C_{d^4D^2}^{(1)}-\frac{2}{3} g_s^2 C_{q^4D^2}^{(1)}-\frac{28}{3} g_s^2 C_{q^2d^2D^2}^{(1)}-2 g_s^2 C_{q^4D^2}^{(3)}+\frac{14}{27} g_1^2 C_{q^2d^2D^2}^{(3)}-4 g_s^2 N_c C_{q^2d^2D^2}^{(3)}+\frac{28 g_s^2 C_{q^2d^2D^2}^{(3)}}{3 N_c}+6 g_s^2 C_{q^2d^2D^2}^{(3)}+2 g_s^2 C_{u^2d^2D^2}^{(3)}+2 g_s^2 C_{q^2u^2D^2}^{(3)}+\frac{14}{3} g_s^2 C_{d^4D^2}^{(2)}-\frac{10}{27} g_1^2 C_{q^2d^2D^2}^{(4)}+4 g_s^2 N_c C_{q^2d^2D^2}^{(4)}-\frac{16 g_s^2 C_{q^2d^2D^2}^{(4)}}{N_c}-3 g_s^2 C_{q^2d^2D^2}^{(4)}+g_s^2 C_{u^2d^2D^2}^{(4)}+2 g_W^2 C_{q^2d^2D^2}^{(4)}+\frac{14}{3} g_s^2 C_{q^4D^2}^{(2)}+14 g_s^2 C_{q^4D^2}^{(4)}-g_s^2 C_{q^2u^2D^2}^{(4)}+\frac{40}{3} g_s^2 C_{q^2d^2D^2}^{(2)}
	\end{dmath*}
	\begin{dmath*}
		\dot{\bar{C}}_{q^2d^2D^2}^{(4)}= \frac{16}{3} g_s^2 C_{q^2d^2D^2}^{(1)}-\frac{8}{27} g_1^2 C_{q^2d^2D^2}^{(3)}+2 g_s^2 N_c C_{q^2d^2D^2}^{(3)}-\frac{16 g_s^2 C_{q^2d^2D^2}^{(3)}}{3 N_c}+\frac{19}{27} g_1^2 C_{q^2d^2D^2}^{(4)}-6 g_s^2 N_c C_{q^2d^2D^2}^{(4)}+\frac{8 g_s^2 C_{q^2d^2D^2}^{(4)}}{N_c}+g_W^2 C_{q^2d^2D^2}^{(4)}-\frac{28}{3} g_s^2 C_{q^2d^2D^2}^{(2)}
	\end{dmath*}
	\begin{dmath*}
		\dot{\bar{C}}_{q^2e^2D^2}^{(1)}= -\frac{8}{9} g_1^2 N_c C_{q^4D^2}^{(1)'}+\frac{1}{3} g_1^2 C_{q^4D^2}^{(1)'}+g_1^2 C_{q^4D^2}^{(3)'}-\frac{4}{9} g_1^2 N_c C_{q^4D^2}^{(2)'}-\frac{7}{3} g_1^2 C_{q^4D^2}^{(2)'}-7 g_1^2 C_{q^4D^2}^{(4)'}-g_1^2 C_{e^4D^2}^{(1)}-\frac{8}{3} g_1^2 N_c C_{q^4D^2}^{(1)}+\frac{1}{9} g_1^2 C_{q^4D^2}^{(1)}-\frac{2}{9} g_1^2 N_c C_{e^2d^2D^2}^{(1)}+\frac{4}{3} g_1^2 N_c C_{q^2d^2D^2}^{(1)}-\frac{2}{3} g_1^2 C_{l^2e^2D^2}^{(1)}+\frac{2}{9} g_1^2 N_c C_{q^2e^2D^2}^{(1)}+\frac{4}{9} g_1^2 N_c C_{e^2u^2D^2}^{(1)}+\frac{50}{9} g_1^2 C_{q^2e^2D^2}^{(1)}+4 g_1^2 C_{l^2q^2D^2}^{(1)}-\frac{8}{3} g_1^2 N_c C_{q^2u^2D^2}^{(1)}+\frac{1}{3} g_1^2 C_{q^4D^2}^{(3)}-\frac{4}{3} g_1^2 N_c C_{q^4D^2}^{(2)}-\frac{7}{9} g_1^2 C_{q^4D^2}^{(2)}-\frac{7}{3} g_1^2 C_{q^4D^2}^{(4)}-\frac{1}{9} g_1^2 N_c C_{e^2d^2D^2}^{(2)}-\frac{2}{3} g_1^2 N_c C_{q^2d^2D^2}^{(2)}+\frac{1}{3} g_1^2 C_{l^2e^2D^2}^{(2)}-\frac{1}{9} g_1^2 N_c C_{q^2e^2D^2}^{(2)}+\frac{2}{9} g_1^2 N_c C_{e^2u^2D^2}^{(2)}-\frac{40}{27} g_1^2 C_{q^2e^2D^2}^{(2)}+\frac{4}{3} g_s^2 N_c C_{q^2e^2D^2}^{(2)}-\frac{4 g_s^2 C_{q^2e^2D^2}^{(2)}}{3 N_c}+2 g_W^2 C_{q^2e^2D^2}^{(2)}+2 g_1^2 C_{l^2q^2D^2}^{(2)}+\frac{4}{3} g_1^2 N_c C_{q^2u^2D^2}^{(2)}
	\end{dmath*}
	\begin{dmath*}
		\dot{\bar{C}}_{q^2e^2D^2}^{(2)}= -\frac{8}{9} g_1^2 C_{q^2e^2D^2}^{(1)}+\frac{79}{27} g_1^2 C_{q^2e^2D^2}^{(2)}+\frac{2}{3} g_s^2 N_c C_{q^2e^2D^2}^{(2)}-\frac{2 g_s^2 C_{q^2e^2D^2}^{(2)}}{3 N_c}+g_W^2 C_{q^2e^2D^2}^{(2)}
	\end{dmath*}
	\begin{dmath*}
		\dot{\bar{C}}_{q^2u^2D^2}^{(1)}= \frac{16}{27} g_1^2 N_c C_{q^4D^2}^{(1)'}+\frac{8}{27} g_1^2 N_c C_{u^4D^2}^{(1)'}-\frac{2}{9} g_1^2 C_{q^4D^2}^{(1)'}-\frac{2}{9} g_1^2 C_{u^4D^2}^{(1)'}-\frac{2}{3} g_1^2 C_{q^4D^2}^{(3)'}+\frac{8}{27} g_1^2 N_c C_{q^4D^2}^{(2)'}+\frac{4}{27} g_1^2 N_c C_{u^4D^2}^{(2)'}+\frac{14}{9} g_1^2 C_{q^4D^2}^{(2)'}+\frac{14}{3} g_1^2 C_{q^4D^2}^{(4)'}+\frac{14}{9} g_1^2 C_{u^4D^2}^{(2)'}+\frac{16}{9} g_1^2 N_c C_{q^4D^2}^{(1)}+\frac{8}{9} g_1^2 N_c C_{u^4D^2}^{(1)}-\frac{2}{27} g_1^2 C_{q^4D^2}^{(1)}-\frac{2}{27} g_1^2 C_{u^4D^2}^{(1)}-\frac{8}{9} g_1^2 N_c C_{q^2d^2D^2}^{(1)}-\frac{2}{9} g_1^2 N_c C_{u^2d^2D^2}^{(1)}-\frac{8}{3} g_1^2 C_{q^2e^2D^2}^{(1)}-\frac{2}{3} g_1^2 C_{e^2u^2D^2}^{(1)}-\frac{8}{3} g_1^2 C_{l^2q^2D^2}^{(1)}-\frac{2}{3} g_1^2 C_{l^2u^2D^2}^{(1)}+2 g_1^2 N_c C_{q^2u^2D^2}^{(1)}-\frac{28}{27} g_1^2 C_{q^2u^2D^2}^{(1)}-\frac{2}{9} g_1^2 C_{q^4D^2}^{(3)}+\frac{7 g_s^2 C_{q^2u^2D^2}^{(3)}}{3 N_c^2}-\frac{7}{3} g_s^2 C_{q^2u^2D^2}^{(3)}+\frac{8}{9} g_1^2 N_c C_{q^4D^2}^{(2)}+\frac{4}{9} g_1^2 N_c C_{u^4D^2}^{(2)}+\frac{14}{27} g_1^2 C_{q^4D^2}^{(2)}+\frac{14}{9} g_1^2 C_{q^4D^2}^{(4)}+\frac{14}{27} g_1^2 C_{u^4D^2}^{(2)}-\frac{10 g_s^2 C_{q^2u^2D^2}^{(4)}}{3 N_c^2}+\frac{10}{3} g_s^2 C_{q^2u^2D^2}^{(4)}+\frac{4}{9} g_1^2 N_c C_{q^2d^2D^2}^{(2)}-\frac{1}{9} g_1^2 N_c C_{u^2d^2D^2}^{(2)}+\frac{4}{3} g_1^2 C_{q^2e^2D^2}^{(2)}-\frac{1}{3} g_1^2 C_{e^2u^2D^2}^{(2)}-\frac{4}{3} g_1^2 C_{l^2q^2D^2}^{(2)}+\frac{1}{3} g_1^2 C_{l^2u^2D^2}^{(2)}-g_1^2 N_c C_{q^2u^2D^2}^{(2)}+\frac{74}{27} g_1^2 C_{q^2u^2D^2}^{(2)}+\frac{8}{3} g_s^2 N_c C_{q^2u^2D^2}^{(2)}-\frac{8 g_s^2 C_{q^2u^2D^2}^{(2)}}{3 N_c}+2 g_W^2 C_{q^2u^2D^2}^{(2)}
	\end{dmath*}
	\begin{dmath*}
		\dot{\bar{C}}_{q^2u^2D^2}^{(2)}= \frac{16}{27} g_1^2 C_{q^2u^2D^2}^{(1)}-\frac{4 g_s^2 C_{q^2u^2D^2}^{(3)}}{3 N_c^2}+\frac{4}{3} g_s^2 C_{q^2u^2D^2}^{(3)}+\frac{7 g_s^2 C_{q^2u^2D^2}^{(4)}}{3 N_c^2}-\frac{7}{3} g_s^2 C_{q^2u^2D^2}^{(4)}-\frac{11}{27} g_1^2 C_{q^2u^2D^2}^{(2)}+\frac{4}{3} g_s^2 N_c C_{q^2u^2D^2}^{(2)}-\frac{4 g_s^2 C_{q^2u^2D^2}^{(2)}}{3 N_c}+g_W^2 C_{q^2u^2D^2}^{(2)}
	\end{dmath*}
	\begin{dmath*}
		\dot{\bar{C}}_{q^2u^2D^2}^{(3)}= -2 g_s^2 C_{q^4D^2}^{(1)'}-2 g_s^2 C_{u^4D^2}^{(1)'}-6 g_s^2 C_{q^4D^2}^{(3)'}+14 g_s^2 C_{q^4D^2}^{(2)'}+42 g_s^2 C_{q^4D^2}^{(4)'}+14 g_s^2 C_{u^4D^2}^{(2)'}-\frac{2}{3} g_s^2 C_{q^4D^2}^{(1)}-\frac{2}{3} g_s^2 C_{u^4D^2}^{(1)}-\frac{28}{3} g_s^2 C_{q^2u^2D^2}^{(1)}-2 g_s^2 C_{q^4D^2}^{(3)}+2 g_s^2 C_{q^2d^2D^2}^{(3)}+2 g_s^2 C_{u^2d^2D^2}^{(3)}-\frac{28}{27} g_1^2 C_{q^2u^2D^2}^{(3)}-4 g_s^2 N_c C_{q^2u^2D^2}^{(3)}+\frac{28 g_s^2 C_{q^2u^2D^2}^{(3)}}{3 N_c}+6 g_s^2 C_{q^2u^2D^2}^{(3)}-g_s^2 C_{q^2d^2D^2}^{(4)}+g_s^2 C_{u^2d^2D^2}^{(4)}+\frac{74}{27} g_1^2 C_{q^2u^2D^2}^{(4)}+4 g_s^2 N_c C_{q^2u^2D^2}^{(4)}-\frac{16 g_s^2 C_{q^2u^2D^2}^{(4)}}{N_c}+\frac{14}{3} g_s^2 C_{q^4D^2}^{(2)}+14 g_s^2 C_{q^4D^2}^{(4)}-3 g_s^2 C_{q^2u^2D^2}^{(4)}+\frac{14}{3} g_s^2 C_{u^4D^2}^{(2)}+2 g_W^2 C_{q^2u^2D^2}^{(4)}+\frac{40}{3} g_s^2 C_{q^2u^2D^2}^{(2)}
	\end{dmath*}
	\begin{dmath*}
		\dot{\bar{C}}_{q^2u^2D^2}^{(4)}= \frac{16}{3} g_s^2 C_{q^2u^2D^2}^{(1)}+\frac{16}{27} g_1^2 C_{q^2u^2D^2}^{(3)}+2 g_s^2 N_c C_{q^2u^2D^2}^{(3)}-\frac{16 g_s^2 C_{q^2u^2D^2}^{(3)}}{3 N_c}-\frac{11}{27} g_1^2 C_{q^2u^2D^2}^{(4)}-6 g_s^2 N_c C_{q^2u^2D^2}^{(4)}+\frac{8 g_s^2 C_{q^2u^2D^2}^{(4)}}{N_c}+g_W^2 C_{q^2u^2D^2}^{(4)}-\frac{28}{3} g_s^2 C_{q^2u^2D^2}^{(2)}
	\end{dmath*}
\end{dgroup}
\subsection{Linear dimension-8 $(\bar RR)(\bar RR)$}
\begin{dgroup}
	\begin{dmath*}
		\dot{\bar{C}}_{e^4D^2}^{(1)}= \frac{94}{3} g_1^2 C_{e^4D^2}^{(1)}+\frac{4}{3} g_1^2 N_c C_{e^2d^2D^2}^{(1)}+4 g_1^2 C_{l^2e^2D^2}^{(1)}-\frac{4}{3} g_1^2 N_c C_{q^2e^2D^2}^{(1)}-\frac{8}{3} g_1^2 N_c C_{e^2u^2D^2}^{(1)}+\frac{2}{3} g_1^2 N_c C_{e^2d^2D^2}^{(2)}-2 g_1^2 C_{l^2e^2D^2}^{(2)}+\frac{2}{3} g_1^2 N_c C_{q^2e^2D^2}^{(2)}-\frac{4}{3} g_1^2 N_c C_{e^2u^2D^2}^{(2)}
	\end{dmath*}
	\begin{dmath*}
		\dot{\bar{C}}_{u^4D^2}^{(1)}= \frac{32}{27} g_1^2 N_c C_{u^4D^2}^{(1)'}-\frac{8}{9} g_1^2 C_{u^4D^2}^{(1)'}+\frac{g_s^2 C_{u^4D^2}^{(1)'}}{N_c}+\frac{5}{3} g_s^2 C_{u^4D^2}^{(1)'}+\frac{16}{27} g_1^2 N_c C_{u^4D^2}^{(2)'}+\frac{56}{9} g_1^2 C_{u^4D^2}^{(2)'}-\frac{7 g_s^2 C_{u^4D^2}^{(2)'}}{N_c}+\frac{11}{3} g_s^2 C_{u^4D^2}^{(2)'}+\frac{32}{9} g_1^2 N_c C_{u^4D^2}^{(1)}+\frac{104}{27} g_1^2 C_{u^4D^2}^{(1)}-\frac{13 g_s^2 C_{u^4D^2}^{(1)}}{3 N_c}-\frac{8}{9} g_1^2 N_c C_{u^2d^2D^2}^{(1)}-\frac{8}{3} g_1^2 C_{e^2u^2D^2}^{(1)}-\frac{8}{3} g_1^2 C_{l^2u^2D^2}^{(1)}+\frac{8}{9} g_1^2 N_c C_{q^2u^2D^2}^{(1)}-\frac{g_s^2 C_{u^2d^2D^2}^{(3)}}{N_c}-\frac{2 g_s^2 C_{q^2u^2D^2}^{(3)}}{N_c}-\frac{g_s^2 C_{u^2d^2D^2}^{(4)}}{2 N_c}+\frac{16}{9} g_1^2 N_c C_{u^4D^2}^{(2)}+\frac{152}{27} g_1^2 C_{u^4D^2}^{(2)}+\frac{g_s^2 C_{q^2u^2D^2}^{(4)}}{N_c}-\frac{8}{3} g_s^2 N_c C_{u^4D^2}^{(2)}-\frac{19 g_s^2 C_{u^4D^2}^{(2)}}{3 N_c}-\frac{4}{9} g_1^2 N_c C_{u^2d^2D^2}^{(2)}-\frac{4}{3} g_1^2 C_{e^2u^2D^2}^{(2)}+\frac{4}{3} g_1^2 C_{l^2u^2D^2}^{(2)}-\frac{4}{9} g_1^2 N_c C_{q^2u^2D^2}^{(2)}
	\end{dmath*}
	\begin{dmath*}
		\dot{\bar{C}}_{u^4D^2}^{(1)'}= \frac{112}{27} g_1^2 C_{u^4D^2}^{(1)'}-\frac{14 g_s^2 C_{u^4D^2}^{(1)'}}{3 N_c}+\frac{1}{2} g_s^2 C_{u^4D^2}^{(1)'}+\frac{32}{9} g_1^2 C_{u^4D^2}^{(2)'}-\frac{8}{3} g_s^2 N_c C_{u^4D^2}^{(2)'}-\frac{4 g_s^2 C_{u^4D^2}^{(2)'}}{N_c}-\frac{7}{2} g_s^2 C_{u^4D^2}^{(2)'}+\frac{11}{6} g_s^2 C_{u^4D^2}^{(1)}-\frac{1}{2} g_s^2 C_{u^2d^2D^2}^{(3)}-g_s^2 C_{q^2u^2D^2}^{(3)}-\frac{1}{4} g_s^2 C_{u^2d^2D^2}^{(4)}+\frac{1}{2} g_s^2 C_{q^2u^2D^2}^{(4)}+\frac{5}{2} g_s^2 C_{u^4D^2}^{(2)}
	\end{dmath*}
	\begin{dmath*}
		\dot{\bar{C}}_{u^4D^2}^{(2)}= \frac{11}{3} g_s^2 C_{u^4D^2}^{(1)'}+\frac{17}{3} g_s^2 C_{u^4D^2}^{(2)'}+\frac{64}{27} g_1^2 C_{u^4D^2}^{(1)}-\frac{8 g_s^2 C_{u^4D^2}^{(1)}}{3 N_c}+\frac{16}{3} g_1^2 C_{u^4D^2}^{(2)}+\frac{4}{3} g_s^2 N_c C_{u^4D^2}^{(2)}-\frac{6 g_s^2 C_{u^4D^2}^{(2)}}{N_c}
	\end{dmath*}
	\begin{dmath*}
		\dot{\bar{C}}_{u^4D^2}^{(2)'}= \frac{64}{27} g_1^2 C_{u^4D^2}^{(1)'}-\frac{8 g_s^2 C_{u^4D^2}^{(1)'}}{3 N_c}-\frac{1}{2} g_s^2 C_{u^4D^2}^{(1)'}+\frac{16}{3} g_1^2 C_{u^4D^2}^{(2)'}+\frac{4}{3} g_s^2 N_c C_{u^4D^2}^{(2)'}-\frac{6 g_s^2 C_{u^4D^2}^{(2)'}}{N_c}+\frac{7}{2} g_s^2 C_{u^4D^2}^{(2)'}+\frac{7}{2} g_s^2 C_{u^4D^2}^{(1)}+\frac{1}{2} g_s^2 C_{u^2d^2D^2}^{(3)}+g_s^2 C_{q^2u^2D^2}^{(3)}+\frac{1}{4} g_s^2 C_{u^2d^2D^2}^{(4)}-\frac{1}{2} g_s^2 C_{q^2u^2D^2}^{(4)}+\frac{41}{6} g_s^2 C_{u^4D^2}^{(2)}
	\end{dmath*}
	\begin{dmath*}
		\dot{\bar{C}}_{d^4D^2}^{(1)}= \frac{8}{27} g_1^2 N_c C_{d^4D^2}^{(1)'}-\frac{2}{9} g_1^2 C_{d^4D^2}^{(1)'}+\frac{g_s^2 C_{d^4D^2}^{(1)'}}{N_c}+\frac{5}{3} g_s^2 C_{d^4D^2}^{(1)'}+\frac{4}{27} g_1^2 N_c C_{d^4D^2}^{(2)'}+\frac{14}{9} g_1^2 C_{d^4D^2}^{(2)'}-\frac{7 g_s^2 C_{d^4D^2}^{(2)'}}{N_c}+\frac{11}{3} g_s^2 C_{d^4D^2}^{(2)'}+\frac{8}{9} g_1^2 N_c C_{d^4D^2}^{(1)}+\frac{26}{27} g_1^2 C_{d^4D^2}^{(1)}-\frac{13 g_s^2 C_{d^4D^2}^{(1)}}{3 N_c}+\frac{4}{3} g_1^2 C_{e^2d^2D^2}^{(1)}+\frac{4}{3} g_1^2 C_{l^2d^2D^2}^{(1)}-\frac{4}{9} g_1^2 N_c C_{q^2d^2D^2}^{(1)}-\frac{8}{9} g_1^2 N_c C_{u^2d^2D^2}^{(1)}-\frac{2 g_s^2 C_{q^2d^2D^2}^{(3)}}{N_c}-\frac{g_s^2 C_{u^2d^2D^2}^{(3)}}{N_c}-\frac{16}{3} C_F g_s^2 C_{d^4D^2}^{(2)}+\frac{4}{9} g_1^2 N_c C_{d^4D^2}^{(2)}+\frac{38}{27} g_1^2 C_{d^4D^2}^{(2)}-\frac{9 g_s^2 C_{d^4D^2}^{(2)}}{N_c}+\frac{g_s^2 C_{q^2d^2D^2}^{(4)}}{N_c}-\frac{g_s^2 C_{u^2d^2D^2}^{(4)}}{2 N_c}+\frac{2}{3} g_1^2 C_{e^2d^2D^2}^{(2)}-\frac{2}{3} g_1^2 C_{l^2d^2D^2}^{(2)}+\frac{2}{9} g_1^2 N_c C_{q^2d^2D^2}^{(2)}-\frac{4}{9} g_1^2 N_c C_{u^2d^2D^2}^{(2)}
	\end{dmath*}
	\begin{dmath*}
		\dot{\bar{C}}_{d^4D^2}^{(1)'}= \frac{28}{27} g_1^2 C_{d^4D^2}^{(1)'}-\frac{14 g_s^2 C_{d^4D^2}^{(1)'}}{3 N_c}+\frac{1}{2} g_s^2 C_{d^4D^2}^{(1)'}-\frac{16}{3} C_F g_s^2 C_{d^4D^2}^{(2)'}+\frac{8}{9} g_1^2 C_{d^4D^2}^{(2)'}-\frac{20 g_s^2 C_{d^4D^2}^{(2)'}}{3 N_c}-\frac{7}{2} g_s^2 C_{d^4D^2}^{(2)'}+\frac{11}{6} g_s^2 C_{d^4D^2}^{(1)}-g_s^2 C_{q^2d^2D^2}^{(3)}-\frac{1}{2} g_s^2 C_{u^2d^2D^2}^{(3)}+\frac{5}{2} g_s^2 C_{d^4D^2}^{(2)}+\frac{1}{2} g_s^2 C_{q^2d^2D^2}^{(4)}-\frac{1}{4} g_s^2 C_{u^2d^2D^2}^{(4)}
	\end{dmath*}
	\begin{dmath*}
		\dot{\bar{C}}_{d^4D^2}^{(2)}= \frac{11}{3} g_s^2 C_{d^4D^2}^{(1)'}+\frac{17}{3} g_s^2 C_{d^4D^2}^{(2)'}+\frac{16}{27} g_1^2 C_{d^4D^2}^{(1)}-\frac{8 g_s^2 C_{d^4D^2}^{(1)}}{3 N_c}+\frac{8}{3} C_F g_s^2 C_{d^4D^2}^{(2)}+\frac{4}{3} g_1^2 C_{d^4D^2}^{(2)}-\frac{14 g_s^2 C_{d^4D^2}^{(2)}}{3 N_c}
	\end{dmath*}
	\begin{dmath*}
		\dot{\bar{C}}_{d^4D^2}^{(2)'}= \frac{16}{27} g_1^2 C_{d^4D^2}^{(1)'}-\frac{8 g_s^2 C_{d^4D^2}^{(1)'}}{3 N_c}-\frac{1}{2} g_s^2 C_{d^4D^2}^{(1)'}+\frac{8}{3} C_F g_s^2 C_{d^4D^2}^{(2)'}+\frac{4}{3} g_1^2 C_{d^4D^2}^{(2)'}-\frac{14 g_s^2 C_{d^4D^2}^{(2)'}}{3 N_c}+\frac{7}{2} g_s^2 C_{d^4D^2}^{(2)'}+\frac{7}{2} g_s^2 C_{d^4D^2}^{(1)}+g_s^2 C_{q^2d^2D^2}^{(3)}+\frac{1}{2} g_s^2 C_{u^2d^2D^2}^{(3)}+\frac{41}{6} g_s^2 C_{d^4D^2}^{(2)}-\frac{1}{2} g_s^2 C_{q^2d^2D^2}^{(4)}+\frac{1}{4} g_s^2 C_{u^2d^2D^2}^{(4)}
	\end{dmath*}
	\begin{dmath*}
		\dot{\bar{C}}_{e^2u^2D^2}^{(1)}= -\frac{16}{9} g_1^2 N_c C_{u^4D^2}^{(1)'}+\frac{4}{3} g_1^2 C_{u^4D^2}^{(1)'}-\frac{8}{9} g_1^2 N_c C_{u^4D^2}^{(2)'}-\frac{28}{3} g_1^2 C_{u^4D^2}^{(2)'}-4 g_1^2 C_{e^4D^2}^{(1)}-\frac{16}{3} g_1^2 N_c C_{u^4D^2}^{(1)}+\frac{4}{9} g_1^2 C_{u^4D^2}^{(1)}-\frac{8}{9} g_1^2 N_c C_{e^2d^2D^2}^{(1)}+\frac{4}{3} g_1^2 N_c C_{u^2d^2D^2}^{(1)}-\frac{8}{3} g_1^2 C_{l^2e^2D^2}^{(1)}+\frac{8}{9} g_1^2 N_c C_{q^2e^2D^2}^{(1)}+\frac{16}{9} g_1^2 N_c C_{e^2u^2D^2}^{(1)}-\frac{20}{9} g_1^2 C_{e^2u^2D^2}^{(1)}+4 g_1^2 C_{l^2u^2D^2}^{(1)}-\frac{4}{3} g_1^2 N_c C_{q^2u^2D^2}^{(1)}-\frac{8}{3} g_1^2 N_c C_{u^4D^2}^{(2)}-\frac{28}{9} g_1^2 C_{u^4D^2}^{(2)}-\frac{4}{9} g_1^2 N_c C_{e^2d^2D^2}^{(2)}+\frac{2}{3} g_1^2 N_c C_{u^2d^2D^2}^{(2)}+\frac{4}{3} g_1^2 C_{l^2e^2D^2}^{(2)}-\frac{4}{9} g_1^2 N_c C_{q^2e^2D^2}^{(2)}+\frac{8}{9} g_1^2 N_c C_{e^2u^2D^2}^{(2)}-\frac{290}{27} g_1^2 C_{e^2u^2D^2}^{(2)}-\frac{4}{3} g_s^2 N_c C_{e^2u^2D^2}^{(2)}+\frac{4 g_s^2 C_{e^2u^2D^2}^{(2)}}{3 N_c}-2 g_1^2 C_{l^2u^2D^2}^{(2)}+\frac{2}{3} g_1^2 N_c C_{q^2u^2D^2}^{(2)}
	\end{dmath*}
	\begin{dmath*}
		\dot{\bar{C}}_{e^2u^2D^2}^{(2)}= -\frac{32}{9} g_1^2 C_{e^2u^2D^2}^{(1)}-\frac{116}{27} g_1^2 C_{e^2u^2D^2}^{(2)}+\frac{2}{3} g_s^2 N_c C_{e^2u^2D^2}^{(2)}-\frac{2 g_s^2 C_{e^2u^2D^2}^{(2)}}{3 N_c}
	\end{dmath*}
	\begin{dmath*}
		\dot{\bar{C}}_{e^2d^2D^2}^{(1)}= \frac{8}{9} g_1^2 N_c C_{d^4D^2}^{(1)'}-\frac{2}{3} g_1^2 C_{d^4D^2}^{(1)'}+\frac{4}{9} g_1^2 N_c C_{d^4D^2}^{(2)'}+\frac{14}{3} g_1^2 C_{d^4D^2}^{(2)'}+\frac{8}{3} g_1^2 N_c C_{d^4D^2}^{(1)}-\frac{2}{9} g_1^2 C_{d^4D^2}^{(1)}+2 g_1^2 C_{e^4D^2}^{(1)}+\frac{4}{9} g_1^2 N_c C_{e^2d^2D^2}^{(1)}+\frac{64}{9} g_1^2 C_{e^2d^2D^2}^{(1)}+4 g_1^2 C_{l^2d^2D^2}^{(1)}-\frac{4}{3} g_1^2 N_c C_{q^2d^2D^2}^{(1)}-\frac{8}{3} g_1^2 N_c C_{u^2d^2D^2}^{(1)}+\frac{4}{3} g_1^2 C_{l^2e^2D^2}^{(1)}-\frac{4}{9} g_1^2 N_c C_{q^2e^2D^2}^{(1)}-\frac{8}{9} g_1^2 N_c C_{e^2u^2D^2}^{(1)}+\frac{4}{3} g_1^2 N_c C_{d^4D^2}^{(2)}+\frac{14}{9} g_1^2 C_{d^4D^2}^{(2)}+\frac{2}{9} g_1^2 N_c C_{e^2d^2D^2}^{(2)}+\frac{94}{27} g_1^2 C_{e^2d^2D^2}^{(2)}-\frac{4}{3} g_s^2 N_c C_{e^2d^2D^2}^{(2)}+\frac{4 g_s^2 C_{e^2d^2D^2}^{(2)}}{3 N_c}-2 g_1^2 C_{l^2d^2D^2}^{(2)}+\frac{2}{3} g_1^2 N_c C_{q^2d^2D^2}^{(2)}-\frac{4}{3} g_1^2 N_c C_{u^2d^2D^2}^{(2)}-\frac{2}{3} g_1^2 C_{l^2e^2D^2}^{(2)}+\frac{2}{9} g_1^2 N_c C_{q^2e^2D^2}^{(2)}-\frac{4}{9} g_1^2 N_c C_{e^2u^2D^2}^{(2)}
	\end{dmath*}
	\begin{dmath*}
		\dot{\bar{C}}_{e^2d^2D^2}^{(2)}= \frac{16}{9} g_1^2 C_{e^2d^2D^2}^{(1)}+\frac{124}{27} g_1^2 C_{e^2d^2D^2}^{(2)}+\frac{2}{3} g_s^2 N_c C_{e^2d^2D^2}^{(2)}-\frac{2 g_s^2 C_{e^2d^2D^2}^{(2)}}{3 N_c}
	\end{dmath*}
	\begin{dmath*}
		\dot{\bar{C}}_{u^2d^2D^2}^{(1)}= -\frac{16}{27} g_1^2 N_c C_{d^4D^2}^{(1)'}+\frac{4}{9} g_1^2 C_{d^4D^2}^{(1)'}-\frac{16}{27} g_1^2 N_c C_{u^4D^2}^{(1)'}+\frac{4}{9} g_1^2 C_{u^4D^2}^{(1)'}-\frac{8}{27} g_1^2 N_c C_{d^4D^2}^{(2)'}-\frac{28}{9} g_1^2 C_{d^4D^2}^{(2)'}-\frac{8}{27} g_1^2 N_c C_{u^4D^2}^{(2)'}-\frac{28}{9} g_1^2 C_{u^4D^2}^{(2)'}-\frac{16}{9} g_1^2 N_c C_{d^4D^2}^{(1)}+\frac{4}{27} g_1^2 C_{d^4D^2}^{(1)}-\frac{16}{9} g_1^2 N_c C_{u^4D^2}^{(1)}+\frac{4}{27} g_1^2 C_{u^4D^2}^{(1)}-\frac{8}{3} g_1^2 C_{e^2d^2D^2}^{(1)}-\frac{8}{3} g_1^2 C_{l^2d^2D^2}^{(1)}+\frac{8}{9} g_1^2 N_c C_{q^2d^2D^2}^{(1)}+\frac{20}{9} g_1^2 N_c C_{u^2d^2D^2}^{(1)}-\frac{56}{27} g_1^2 C_{u^2d^2D^2}^{(1)}+\frac{4}{3} g_1^2 C_{e^2u^2D^2}^{(1)}+\frac{4}{3} g_1^2 C_{l^2u^2D^2}^{(1)}-\frac{4}{9} g_1^2 N_c C_{q^2u^2D^2}^{(1)}-\frac{7 g_s^2 C_{u^2d^2D^2}^{(3)}}{3 N_c^2}+\frac{7}{3} g_s^2 C_{u^2d^2D^2}^{(3)}-\frac{8}{9} g_1^2 N_c C_{d^4D^2}^{(2)}-\frac{28}{27} g_1^2 C_{d^4D^2}^{(2)}-\frac{10 g_s^2 C_{u^2d^2D^2}^{(4)}}{3 N_c^2}+\frac{10}{3} g_s^2 C_{u^2d^2D^2}^{(4)}-\frac{8}{9} g_1^2 N_c C_{u^4D^2}^{(2)}-\frac{28}{27} g_1^2 C_{u^4D^2}^{(2)}-\frac{4}{3} g_1^2 C_{e^2d^2D^2}^{(2)}+\frac{4}{3} g_1^2 C_{l^2d^2D^2}^{(2)}-\frac{4}{9} g_1^2 N_c C_{q^2d^2D^2}^{(2)}+\frac{10}{9} g_1^2 N_c C_{u^2d^2D^2}^{(2)}-\frac{40}{9} g_1^2 C_{u^2d^2D^2}^{(2)}-\frac{8}{3} g_s^2 N_c C_{u^2d^2D^2}^{(2)}+\frac{8 g_s^2 C_{u^2d^2D^2}^{(2)}}{3 N_c}+\frac{2}{3} g_1^2 C_{e^2u^2D^2}^{(2)}-\frac{2}{3} g_1^2 C_{l^2u^2D^2}^{(2)}+\frac{2}{9} g_1^2 N_c C_{q^2u^2D^2}^{(2)}
	\end{dmath*}
	\begin{dmath*}
		\dot{\bar{C}}_{u^2d^2D^2}^{(2)}= -\frac{32}{27} g_1^2 C_{u^2d^2D^2}^{(1)}-\frac{4 g_s^2 C_{u^2d^2D^2}^{(3)}}{3 N_c^2}+\frac{4}{3} g_s^2 C_{u^2d^2D^2}^{(3)}-\frac{7 g_s^2 C_{u^2d^2D^2}^{(4)}}{3 N_c^2}+\frac{7}{3} g_s^2 C_{u^2d^2D^2}^{(4)}-\frac{4}{3} g_1^2 C_{u^2d^2D^2}^{(2)}+\frac{4}{3} g_s^2 N_c C_{u^2d^2D^2}^{(2)}-\frac{4 g_s^2 C_{u^2d^2D^2}^{(2)}}{3 N_c}
	\end{dmath*}
	\begin{dmath*}
		\dot{\bar{C}}_{u^2d^2D^2}^{(3)}= -2 g_s^2 C_{d^4D^2}^{(1)'}-2 g_s^2 C_{u^4D^2}^{(1)'}+14 g_s^2 C_{d^4D^2}^{(2)'}+14 g_s^2 C_{u^4D^2}^{(2)'}-\frac{2}{3} g_s^2 C_{d^4D^2}^{(1)}-\frac{2}{3} g_s^2 C_{u^4D^2}^{(1)}+\frac{28}{3} g_s^2 C_{u^2d^2D^2}^{(1)}-\frac{56}{27} g_1^2 C_{u^2d^2D^2}^{(3)}+\frac{2}{3} g_s^2 N_c C_{u^2d^2D^2}^{(3)}-\frac{28 g_s^2 C_{u^2d^2D^2}^{(3)}}{3 N_c}+4 g_s^2 C_{q^2d^2D^2}^{(3)}+4 g_s^2 C_{u^2d^2D^2}^{(3)}+4 g_s^2 C_{q^2u^2D^2}^{(3)}+\frac{14}{3} g_s^2 C_{d^4D^2}^{(2)}-\frac{40}{9} g_1^2 C_{u^2d^2D^2}^{(4)}+\frac{8}{3} g_s^2 N_c C_{u^2d^2D^2}^{(4)}-\frac{32 g_s^2 C_{u^2d^2D^2}^{(4)}}{3 N_c}-2 g_s^2 C_{q^2d^2D^2}^{(4)}+2 g_s^2 C_{u^2d^2D^2}^{(4)}-2 g_s^2 C_{q^2u^2D^2}^{(4)}+\frac{14}{3} g_s^2 C_{u^4D^2}^{(2)}+\frac{40}{3} g_s^2 C_{u^2d^2D^2}^{(2)}
	\end{dmath*}
	\begin{dmath*}
		\dot{\bar{C}}_{u^2d^2D^2}^{(4)}= \frac{16}{3} g_s^2 C_{u^2d^2D^2}^{(1)}-\frac{32}{27} g_1^2 C_{u^2d^2D^2}^{(3)}+\frac{2}{3} g_s^2 N_c C_{u^2d^2D^2}^{(3)}-\frac{16 g_s^2 C_{u^2d^2D^2}^{(3)}}{3 N_c}-\frac{4}{3} g_1^2 C_{u^2d^2D^2}^{(4)}-\frac{4}{3} g_s^2 N_c C_{u^2d^2D^2}^{(4)}-\frac{32 g_s^2 C_{u^2d^2D^2}^{(4)}}{3 N_c}+\frac{28}{3} g_s^2 C_{u^2d^2D^2}^{(2)}
	\end{dmath*}
\end{dgroup}

\allowdisplaybreaks
\subsection{Quadratic dimension-6 $(\bar LL)(\bar LL)$}
\begin{align*}
	&\begin{aligned}
		\dot{\hat{C}}_{l^4D^2}^{(1)}=& 4 N_c {C_{lq}^{(1)}}^2-4 N_c {C_{lq}^{(3)}}^2+\frac{8}{3} \left(C_{ll}^'\right)^2+\frac{64}{3} C_{ll} C_{ll}^'+2 N_c C_{ld}^2+2 C_{le}^2+2 N_c C_{lu}^2+32 C_{ll}^2
	\end{aligned}\\
	&\begin{aligned}
		\dot{\hat{C}}_{l^4D^2}^{(1)'}=& -4 N_c {C_{lq}^{(3)}}^2-\frac{8}{3} C_{ll} C_{ll}^'-4 C_{ll}^2
	\end{aligned}\\
	&\begin{aligned}
		\dot{\hat{C}}_{l^4D^2}^{(2)}=& \frac{16}{3} \left(C_{ll}^'\right)^2
	\end{aligned}\\
	&\begin{aligned}
		\dot{\hat{C}}_{l^4D^2}^{(2)'}=& 4 N_c {C_{lq}^{(3)}}^2+\frac{8}{3} C_{ll} C_{ll}^'+4 C_{ll}^2
	\end{aligned}\\
	&\begin{aligned}
		\dot{\hat{C}}_{q^4D^2}^{(1)}=& 48 C_{qq}^{(1)} C_{qq}^{(3)'}+16 C_{qq}^{(1)'} C_{qq}^{(3)'}+\frac{32}{3} N_c C_{qq}^{(1)} C_{qq}^{(1)'}+\frac{28}{3} \left(C_{qq}^{(1)'}\right)^2+16 C_{qq}^{(1)} C_{qq}^{(1)'}+12 \left(C_{qq}^{(3)'}\right)^2
		\\&+16 C_{qq}^{(1)} C_{qq}^{(3)}+2 N_c {C_{qd}^{(1)}}^2+4 {C_{lq}^{(1)}}^2+2 N_c {C_{qu}^{(1)}}^2+16 N_c {C_{qq}^{(1)}}^2+8 {C_{qq}^{(1)}}^2+8 {C_{qq}^{(3)}}^2-\frac{{C_{qd}^{(8)}}^2}{2 N_c}
		\\&-\frac{{C_{qu}^{(8)}}^2}{2 N_c}+2 C_{qe}^2
	\end{aligned}\\
	&\begin{aligned}
		\dot{\hat{C}}_{q^4D^2}^{(1)'}=& 16 C_{qq}^{(3)} C_{qq}^{(1)'}+\frac{16}{3} N_c \left(C_{qq}^{(1)'}\right)^2-4 \left(C_{qq}^{(1)'}\right)^2+\frac{32}{3} C_{qq}^{(1)} C_{qq}^{(1)'}-12 \left(C_{qq}^{(3)'}\right)^2+16 C_{qq}^{(3)} C_{qq}^{(3)'}
		\\&-\frac{1}{8} {C_{qd}^{(8)}}^2-\frac{1}{8} {C_{qu}^{(8)}}^2
	\end{aligned}\\
	&\begin{aligned}
		\dot{\hat{C}}_{q^4D^2}^{(2)}=& 4 \left(C_{qq}^{(1)'}\right)^2+12 \left(C_{qq}^{(3)'}\right)^2+\frac{16}{3} {C_{qq}^{(1)}}^2+16 {C_{qq}^{(3)}}^2
	\end{aligned}\\
	&\begin{aligned}
		\dot{\hat{C}}_{q^4D^2}^{(2)'}=& 4 \left(C_{qq}^{(1)'}\right)^2+\frac{32}{3} C_{qq}^{(1)} C_{qq}^{(1)'}+12 \left(C_{qq}^{(3)'}\right)^2+32 C_{qq}^{(3)} C_{qq}^{(3)'}+\frac{1}{8} {C_{qd}^{(8)}}^2+\frac{1}{8} {C_{qu}^{(8)}}^2
	\end{aligned}\\
	&\begin{aligned}
		\dot{\hat{C}}_{q^4D^2}^{(3)}=& 16 C_{qq}^{(3)} C_{qq}^{(1)'}+\frac{40}{3} C_{qq}^{(1)'} C_{qq}^{(3)'}+\frac{32}{3} N_c C_{qq}^{(3)} C_{qq}^{(3)'}-\frac{40}{3} \left(C_{qq}^{(3)'}\right)^2-16 C_{qq}^{(3)} C_{qq}^{(3)'}+\frac{32}{3} C_{qq}^{(1)} C_{qq}^{(3)}
		\\&+4 {C_{lq}^{(3)}}^2+16 N_c {C_{qq}^{(3)}}^2-16 {C_{qq}^{(3)}}^2
	\end{aligned}\\
	&\begin{aligned}
		\dot{\hat{C}}_{q^4D^2}^{(3)'}=& \frac{16}{3} C_{qq}^{(3)} C_{qq}^{(1)'}+\frac{32}{3} C_{qq}^{(1)} C_{qq}^{(3)'}-8 C_{qq}^{(1)'} C_{qq}^{(3)'}+\frac{16}{3} N_c \left(C_{qq}^{(3)'}\right)^2-8 \left(C_{qq}^{(3)'}\right)^2-\frac{80}{3} C_{qq}^{(3)} C_{qq}^{(3)'}
		\\&-\frac{1}{8} {C_{qd}^{(8)}}^2-\frac{1}{8} {C_{qu}^{(8)}}^2
	\end{aligned}\\
	&\begin{aligned}
		\dot{\hat{C}}_{q^4D^2}^{(4)}=& 8 C_{qq}^{(1)'} C_{qq}^{(3)'}-8 \left(C_{qq}^{(3)'}\right)^2+\frac{32}{3} C_{qq}^{(1)} C_{qq}^{(3)}-\frac{16}{3} {C_{qq}^{(3)}}^2
	\end{aligned}\\
	&\begin{aligned}
		\dot{\hat{C}}_{q^4D^2}^{(4)'}=& \frac{32}{3} C_{qq}^{(3)} C_{qq}^{(1)'}+\frac{32}{3} C_{qq}^{(1)} C_{qq}^{(3)'}+8 C_{qq}^{(1)'} C_{qq}^{(3)'}+8 \left(C_{qq}^{(3)'}\right)^2-\frac{32}{3} C_{qq}^{(3)} C_{qq}^{(3)'}+\frac{1}{8} {C_{qd}^{(8)}}^2+\frac{1}{8} {C_{qu}^{(8)}}^2
	\end{aligned}\\
	&\begin{aligned}
		\dot{\hat{C}}_{l^2q^2D^2}^{(1)}=& 24 C_{lq}^{(1)} C_{qq}^{(3)'}+\frac{16}{3} N_c C_{lq}^{(1)} C_{qq}^{(1)'}+8 C_{ll}^' C_{lq}^{(1)}+8 C_{lq}^{(1)} C_{qq}^{(1)'}+8 C_{qq}^{(3)} C_{lq}^{(1)}+4 N_c C_{ld} C_{qd}^{(1)}
		\\&+4 N_c C_{lu} C_{qu}^{(1)}+16 N_c C_{qq}^{(1)} C_{lq}^{(1)}+\frac{4}{3} {C_{lq}^{(1)}}^2+24 C_{ll} C_{lq}^{(1)}+\frac{8}{3} C_{qq}^{(1)} C_{lq}^{(1)}+4 {C_{lq}^{(3)}}^2+4 C_{le} C_{qe}
	\end{aligned}\\
	&\begin{aligned}
		\dot{\hat{C}}_{l^2q^2D^2}^{(2)}=& \frac{8}{3} {C_{lq}^{(1)}}^2+8 {C_{lq}^{(3)}}^2
	\end{aligned}\\
	&\begin{aligned}
		\dot{\hat{C}}_{l^2q^2D^2}^{(3)}=& 8 C_{lq}^{(3)} C_{qq}^{(1)'}+\frac{16}{3} N_c C_{lq}^{(3)} C_{qq}^{(3)'}+\frac{8}{3} C_{ll}^' C_{lq}^{(3)}-8 C_{lq}^{(3)} C_{qq}^{(3)'}+\frac{8}{3} C_{lq}^{(1)} C_{lq}^{(3)}+\frac{8}{3} C_{qq}^{(1)} C_{lq}^{(3)}
		\\&+16 N_c C_{qq}^{(3)} C_{lq}^{(3)}-\frac{16}{3} {C_{lq}^{(3)}}^2+8 C_{ll} C_{lq}^{(3)}-\frac{8}{3} C_{qq}^{(3)} C_{lq}^{(3)}
	\end{aligned}\\
	&\begin{aligned}
		\dot{\hat{C}}_{l^2q^2D^2}^{(4)}=& \frac{16}{3} C_{lq}^{(1)} C_{lq}^{(3)}-\frac{8}{3} {C_{lq}^{(3)}}^2
	\end{aligned}
\end{align*}
\subsection{Quadratic dimension-6 $(\bar LL)(\bar RR)$}
\begin{align*}
	&\begin{aligned}
		\dot{\hat{C}}_{l^2d^2D^2}^{(1)}=& 8 N_c C_{qd}^{(1)} C_{lq}^{(1)}+4 N_c C_{lu} C_{ud}^{(1)}+\frac{8}{3} N_c C_{ld} C_{dd}^'+8 C_{ld} C_{dd}^'+8 C_{ld} C_{ll}^'+4 C_{ed} C_{le}+8 N_c C_{dd} C_{ld}
		\\&-\frac{4 C_{ld}^2}{3}+\frac{8}{3} C_{dd} C_{ld}+24 C_{ll} C_{ld}
	\end{aligned}\\
	&\begin{aligned}
		\dot{\hat{C}}_{l^2d^2D^2}^{(2)}=& \frac{8 C_{ld}^2}{3}
	\end{aligned}\\
	&\begin{aligned}
		\dot{\hat{C}}_{l^2e^2D^2}^{(1)}=& 8 N_c C_{qe} C_{lq}^{(1)}+8 C_{le} C_{ll}^'+4 N_c C_{ed} C_{ld}+4 N_c C_{eu} C_{lu}-\frac{4 C_{le}^2}{3}+16 C_{ee} C_{le}+24 C_{ll} C_{le}
	\end{aligned}\\
	&\begin{aligned}
		\dot{\hat{C}}_{l^2e^2D^2}^{(2)}=& \frac{8 C_{le}^2}{3}
	\end{aligned}\\
	&\begin{aligned}
		\dot{\hat{C}}_{l^2u^2D^2}^{(1)}=& 4 N_c C_{ld} C_{ud}^{(1)}+8 N_c C_{lq}^{(1)} C_{qu}^{(1)}+\frac{8}{3} N_c C_{lu} C_{uu}^'+8 C_{lu} C_{ll}^'+8 C_{lu} C_{uu}^'+4 C_{le} C_{eu}+8 N_c C_{uu} C_{lu}
		\\&-\frac{4 C_{lu}^2}{3}+24 C_{ll} C_{lu}+\frac{8}{3} C_{uu} C_{lu}
	\end{aligned}\\
	&\begin{aligned}
		\dot{\hat{C}}_{l^2u^2D^2}^{(2)}=& \frac{8 C_{lu}^2}{3}
	\end{aligned}\\
	&\begin{aligned}
		\dot{\hat{C}}_{q^2d^2D^2}^{(1)}=& 24 C_{qd}^{(1)} C_{qq}^{(3)'}+\frac{8}{3} N_c C_{dd}^' C_{qd}^{(1)}+\frac{16}{3} N_c C_{qd}^{(1)} C_{qq}^{(1)'}+8 C_{dd}^' C_{qd}^{(1)}+8 C_{qd}^{(1)} C_{qq}^{(1)'}+8 C_{qq}^{(3)} C_{qd}^{(1)}
		\\&+8 C_{ld} C_{lq}^{(1)}+4 N_c C_{ud}^{(1)} C_{qu}^{(1)}+8 N_c C_{dd} C_{qd}^{(1)}+16 N_c C_{qq}^{(1)} C_{qd}^{(1)}-\frac{4}{3} {C_{qd}^{(1)}}^2+\frac{8}{3} C_{dd} C_{qd}^{(1)}
		\\&+\frac{8}{3} C_{qq}^{(1)} C_{qd}^{(1)}+\frac{{C_{qd}^{(8)}}^2}{3 N_c^2}-\frac{1}{3} {C_{qd}^{(8)}}^2+4 C_{ed} C_{qe}
	\end{aligned}\\
	&\begin{aligned}
		\dot{\hat{C}}_{q^2d^2D^2}^{(2)}=& \frac{8}{3} {C_{qd}^{(1)}}^2-\frac{2 {C_{qd}^{(8)}}^2}{3 N_c^2}+\frac{2}{3} {C_{qd}^{(8)}}^2
	\end{aligned}\\
	&\begin{aligned}
		\dot{\hat{C}}_{q^2d^2D^2}^{(3)}=& 8 C_{qd}^{(8)} C_{qq}^{(1)'}+24 C_{qd}^{(8)} C_{qq}^{(3)'}+8 C_{dd}^' C_{qd}^{(8)}-\frac{8}{3} C_{qd}^{(1)} C_{qd}^{(8)}+\frac{8}{3} C_{qq}^{(1)} C_{qd}^{(8)}+8 C_{qq}^{(3)} C_{qd}^{(8)}-N_c {C_{qd}^{(8)}}^2
		\\&+\frac{4 {C_{qd}^{(8)}}^2}{3 N_c}+2 C_{ud}^{(8)} C_{qu}^{(8)}+\frac{8}{3} C_{dd} C_{qd}^{(8)}
	\end{aligned}\\
	&\begin{aligned}
		\dot{\hat{C}}_{q^2d^2D^2}^{(4)}=& \frac{16}{3} C_{qd}^{(1)} C_{qd}^{(8)}+N_c {C_{qd}^{(8)}}^2-\frac{8 {C_{qd}^{(8)}}^2}{3 N_c}
	\end{aligned}\\
	&\begin{aligned}
		\dot{\hat{C}}_{q^2e^2D^2}^{(1)}=& \frac{16}{3} N_c C_{qe} C_{qq}^{(1)'}+8 C_{qe} C_{qq}^{(1)'}+24 C_{qe} C_{qq}^{(3)'}+4 N_c C_{ed} C_{qd}^{(1)}+8 C_{le} C_{lq}^{(1)}+4 N_c C_{eu} C_{qu}^{(1)}
		\\&+16 N_c C_{qe} C_{qq}^{(1)}+\frac{8}{3} C_{qe} C_{qq}^{(1)}+8 C_{qe} C_{qq}^{(3)}-\frac{4 C_{qe}^2}{3}+16 C_{ee} C_{qe}
	\end{aligned}\\
	&\begin{aligned}
		\dot{\hat{C}}_{q^2e^2D^2}^{(2)}=& \frac{8 C_{qe}^2}{3}
	\end{aligned}\\
	&\begin{aligned}
		\dot{\hat{C}}_{q^2u^2D^2}^{(1)}=& 24 C_{qu}^{(1)} C_{qq}^{(3)'}+\frac{16}{3} N_c C_{qu}^{(1)} C_{qq}^{(1)'}+\frac{8}{3} N_c C_{uu}^' C_{qu}^{(1)}+8 C_{qu}^{(1)} C_{qq}^{(1)'}+8 C_{uu}^' C_{qu}^{(1)}+8 C_{qq}^{(3)} C_{qu}^{(1)}
		\\&+4 N_c C_{qd}^{(1)} C_{ud}^{(1)}+8 C_{lu} C_{lq}^{(1)}+16 N_c C_{qq}^{(1)} C_{qu}^{(1)}+8 N_c C_{uu} C_{qu}^{(1)}-\frac{4}{3} {C_{qu}^{(1)}}^2+\frac{8}{3} C_{qq}^{(1)} C_{qu}^{(1)}
		\\&+\frac{8}{3} C_{uu} C_{qu}^{(1)}+\frac{{C_{qu}^{(8)}}^2}{3 N_c^2}-\frac{1}{3} {C_{qu}^{(8)}}^2+4 C_{qe} C_{eu}
	\end{aligned}\\
	&\begin{aligned}
		\dot{\hat{C}}_{q^2u^2D^2}^{(2)}=& \frac{8}{3} {C_{qu}^{(1)}}^2-\frac{2 {C_{qu}^{(8)}}^2}{3 N_c^2}+\frac{2}{3} {C_{qu}^{(8)}}^2
	\end{aligned}\\
	&\begin{aligned}
		\dot{\hat{C}}_{q^2u^2D^2}^{(3)}=& 8 C_{qu}^{(8)} C_{qq}^{(1)'}+24 C_{qu}^{(8)} C_{qq}^{(3)'}+8 C_{uu}^' C_{qu}^{(8)}+\frac{8}{3} C_{qq}^{(1)} C_{qu}^{(8)}-\frac{8}{3} C_{qu}^{(1)} C_{qu}^{(8)}+8 C_{qq}^{(3)} C_{qu}^{(8)}+2 C_{qd}^{(8)} C_{ud}^{(8)}
		\\&-N_c {C_{qu}^{(8)}}^2+\frac{4 {C_{qu}^{(8)}}^2}{3 N_c}+\frac{8}{3} C_{uu} C_{qu}^{(8)}
	\end{aligned}\\
	&\begin{aligned}
		\dot{\hat{C}}_{q^2u^2D^2}^{(4)}=& \frac{16}{3} C_{qu}^{(1)} C_{qu}^{(8)}+N_c {C_{qu}^{(8)}}^2-\frac{8 {C_{qu}^{(8)}}^2}{3 N_c}
	\end{aligned}\\
\end{align*}
\subsection{Quadratic dimension-6 $(\bar RR)(\bar RR)$}
\begin{align*}
	&\begin{aligned}
		\dot{\hat{C}}_{e^4D^2}^{(1)}=& 2 N_c C_{ed}^2+4 C_{le}^2+4 N_c C_{qe}^2+2 N_c C_{eu}^2+32 C_{ee}^2
	\end{aligned}\\
	&\begin{aligned}
		\dot{\hat{C}}_{u^4D^2}^{(1)}=& 2 N_c {C_{ud}^{(1)}}^2+4 N_c {C_{qu}^{(1)}}^2-\frac{{C_{ud}^{(8)}}^2}{2 N_c}-\frac{{C_{qu}^{(8)}}^2}{N_c}+\frac{16}{3} N_c C_{uu} C_{uu}^'+\frac{28}{3} \left(C_{uu}^'\right)^2+16 C_{uu} C_{uu}^'+2 C_{eu}^2
		\\&+4 C_{lu}^2+8 N_c C_{uu}^2+8 C_{uu}^2
	\end{aligned}\\
	&\begin{aligned}
		\dot{\hat{C}}_{u^4D^2}^{(1)'}=& -\frac{1}{4} {C_{ud}^{(8)}}^2-\frac{1}{2} {C_{qu}^{(8)}}^2+\frac{8}{3} N_c \left(C_{uu}^'\right)^2-4 \left(C_{uu}^'\right)^2+\frac{32}{3} C_{uu} C_{uu}^'
	\end{aligned}\\
	&\begin{aligned}
		\dot{\hat{C}}_{u^4D^2}^{(2)}=& 4 \left(C_{uu}^'\right)^2+\frac{16 C_{uu}^2}{3}
	\end{aligned}\\
	&\begin{aligned}
		\dot{\hat{C}}_{u^4D^2}^{(2)'}=& \frac{1}{4} {C_{ud}^{(8)}}^2+\frac{1}{2} {C_{qu}^{(8)}}^2+4 \left(C_{uu}^'\right)^2+\frac{32}{3} C_{uu} C_{uu}^'
	\end{aligned}\\
	&\begin{aligned}
		\dot{\hat{C}}_{d^4D^2}^{(1)}=& 4 N_c {C_{qd}^{(1)}}^2+2 N_c {C_{ud}^{(1)}}^2-\frac{{C_{qd}^{(8)}}^2}{N_c}-\frac{{C_{ud}^{(8)}}^2}{2 N_c}+\frac{16}{3} N_c C_{dd} C_{dd}^'+16 C_{dd} C_{dd}^'+\frac{28}{3} \left(C_{dd}^'\right)^2+2 C_{ed}^2
		\\&+4 C_{ld}^2+8 N_c C_{dd}^2+8 C_{dd}^2
	\end{aligned}\\
	&\begin{aligned}
		\dot{\hat{C}}_{d^4D^2}^{(1)'}=& -\frac{1}{2} {C_{qd}^{(8)}}^2-\frac{1}{4} {C_{ud}^{(8)}}^2+\frac{8}{3} N_c \left(C_{dd}^'\right)^2-4 \left(C_{dd}^'\right)^2+\frac{32}{3} C_{dd} C_{dd}^'
	\end{aligned}\\
	&\begin{aligned}
		\dot{\hat{C}}_{d^4D^2}^{(2)}=& 4 \left(C_{dd}^'\right)^2+\frac{16 C_{dd}^2}{3}
	\end{aligned}\\
	&\begin{aligned}
		\dot{\hat{C}}_{d^4D^2}^{(2)'}=& \frac{1}{2} {C_{qd}^{(8)}}^2+\frac{1}{4} {C_{ud}^{(8)}}^2+4 \left(C_{dd}^'\right)^2+\frac{32}{3} C_{dd} C_{dd}^'
	\end{aligned}\\
	&\begin{aligned}
		\dot{\hat{C}}_{e^2u^2D^2}^{(1)}=& 4 N_c C_{ed} C_{ud}^{(1)}+8 N_c C_{qe} C_{qu}^{(1)}+\frac{8}{3} N_c C_{eu} C_{uu}^'+8 C_{eu} C_{uu}^'+8 C_{le} C_{lu}+8 N_c C_{uu} C_{eu}+\frac{4 C_{eu}^2}{3}
		\\&+16 C_{ee} C_{eu}+\frac{8}{3} C_{uu} C_{eu}
	\end{aligned}\\
	&\begin{aligned}
		\dot{\hat{C}}_{e^2u^2D^2}^{(2)}=& \frac{8 C_{eu}^2}{3}
	\end{aligned}\\
	&\begin{aligned}
		\dot{\hat{C}}_{e^2d^2D^2}^{(1)}=& 8 N_c C_{qe} C_{qd}^{(1)}+4 N_c C_{eu} C_{ud}^{(1)}+\frac{8}{3} N_c C_{ed} C_{dd}^'+8 C_{ed} C_{dd}^'+8 C_{ld} C_{le}+8 N_c C_{dd} C_{ed}+\frac{4 C_{ed}^2}{3}
		\\&+\frac{8}{3} C_{dd} C_{ed}+16 C_{ee} C_{ed}
	\end{aligned}\\
	&\begin{aligned}
		\dot{\hat{C}}_{e^2d^2D^2}^{(2)}=& \frac{8 C_{ed}^2}{3}
	\end{aligned}\\
	&\begin{aligned}
		\dot{\hat{C}}_{u^2d^2D^2}^{(1)}=& \frac{8}{3} N_c C_{dd}^' C_{ud}^{(1)}+\frac{8}{3} N_c C_{uu}^' C_{ud}^{(1)}+8 C_{dd}^' C_{ud}^{(1)}+8 C_{uu}^' C_{ud}^{(1)}+8 N_c C_{qd}^{(1)} C_{qu}^{(1)}+8 N_c C_{dd} C_{ud}^{(1)}
		\\&+8 N_c C_{uu} C_{ud}^{(1)}+\frac{4}{3} {C_{ud}^{(1)}}^2+\frac{8}{3} C_{dd} C_{ud}^{(1)}+\frac{8}{3} C_{uu} C_{ud}^{(1)}-\frac{{C_{ud}^{(8)}}^2}{3 N_c^2}+\frac{1}{3} {C_{ud}^{(8)}}^2+4 C_{ed} C_{eu}+8 C_{ld} C_{lu}
	\end{aligned}\\
	&\begin{aligned}
		\dot{\hat{C}}_{u^2d^2D^2}^{(2)}=& \frac{8}{3} {C_{ud}^{(1)}}^2-\frac{2 {C_{ud}^{(8)}}^2}{3 N_c^2}+\frac{2}{3} {C_{ud}^{(8)}}^2
	\end{aligned}\\
	&\begin{aligned}
		\dot{\hat{C}}_{u^2d^2D^2}^{(3)}=& 8 C_{dd}^' C_{ud}^{(8)}+8 C_{uu}^' C_{ud}^{(8)}+\frac{8}{3} C_{ud}^{(1)} C_{ud}^{(8)}-\frac{1}{3} N_c {C_{ud}^{(8)}}^2-\frac{4 {C_{ud}^{(8)}}^2}{3 N_c}+4 C_{qd}^{(8)} C_{qu}^{(8)}+\frac{8}{3} C_{dd} C_{ud}^{(8)}+\frac{8}{3} C_{uu} C_{ud}^{(8)}
	\end{aligned}\\
	&\begin{aligned}
		\dot{\hat{C}}_{u^2d^2D^2}^{(4)}=& \frac{16}{3} C_{ud}^{(1)} C_{ud}^{(8)}+\frac{1}{3} N_c {C_{ud}^{(8)}}^2-\frac{8 {C_{ud}^{(8)}}^2}{3 N_c}
	\end{aligned}\\
\end{align*}

\bibliographystyle{h-physrev}
\bibliography{RGE}

\end{document}